\documentclass{sig-alternate}
\usepackage[english]{babel}
\usepackage[latin1]{inputenc}
\usepackage{amssymb}
\usepackage[ruled]{algorithm2e}
\usepackage{algpseudocode}
\usepackage{bm}
\usepackage{amsmath}
\usepackage{graphicx}
\usepackage{epsfig}
\usepackage{url}
\usepackage{psfrag}
\usepackage{balance}
\usepackage{pstricks}
\usepackage{times}
\usepackage{array}
\usepackage{subfigure}
\usepackage{float}
\usepackage{multirow}
\usepackage{mathtools}

\newtheorem{theorem}{\textbf{Theorem}}

\begin{document}

\title{Moss: A Scalable Tool for Efficiently Sampling and Counting 4- and 5-Node Graphlets}
\numberofauthors{1}
\author{
\alignauthor
Pinghui Wang, Jing Tao, Junzhou Zhao, and Xiaohong Guan\\
{\small
\affaddr{MOE Key Laboratory for Intelligent Networks and Network Security, Xi'an Jiaotong University, China}\\
\{phwang, jtao, jzzhao, xhguan\}@sei.xjtu.edu.cn
}
}

\maketitle

\begin{abstract}
Counting the frequencies of 3-, 4-, and 5-node undirected motifs (also know as graphlets)
is widely used for understanding complex networks such as social and biology networks.
However, it is a great challenge to compute these metrics for a large graph due to the intensive computation.
Despite recent efforts to count triangles (i.e., 3-node undirected motif counting),
little attention has been given to developing scalable tools that can be used to characterize 4- and 5-node motifs.
In this paper, we develop computational efficient methods to sample and count 4- and 5- node undirected motifs.
Our methods provide unbiased estimators of motif frequencies, and we derive simple and exact formulas for the variances of the estimators.
Moreover, our methods are designed to fit vertex centric programming models, so they can be easily applied to current graph computing systems such as Pregel and GraphLab.
We conduct experiments on a variety of real-word datasets,
and experimental results show that our methods are several orders of magnitude faster than the state-of-the-art methods under the same estimation errors.
\end{abstract}

\section{Introduction} \label{sec:introduction}
\begin{figure*}[htb]
\center
\subfigure[4-node undirected motifs $M_i^{(4)}, 1\le i\le 6$.]{
\includegraphics[width=0.62\textwidth]{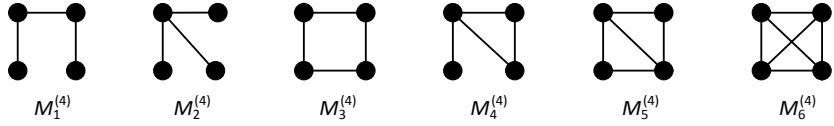}}
\subfigure[5-node undirected motifs $M_i^{(5)}, 1\le i\le 21$.]{
\includegraphics[width=0.99\textwidth]{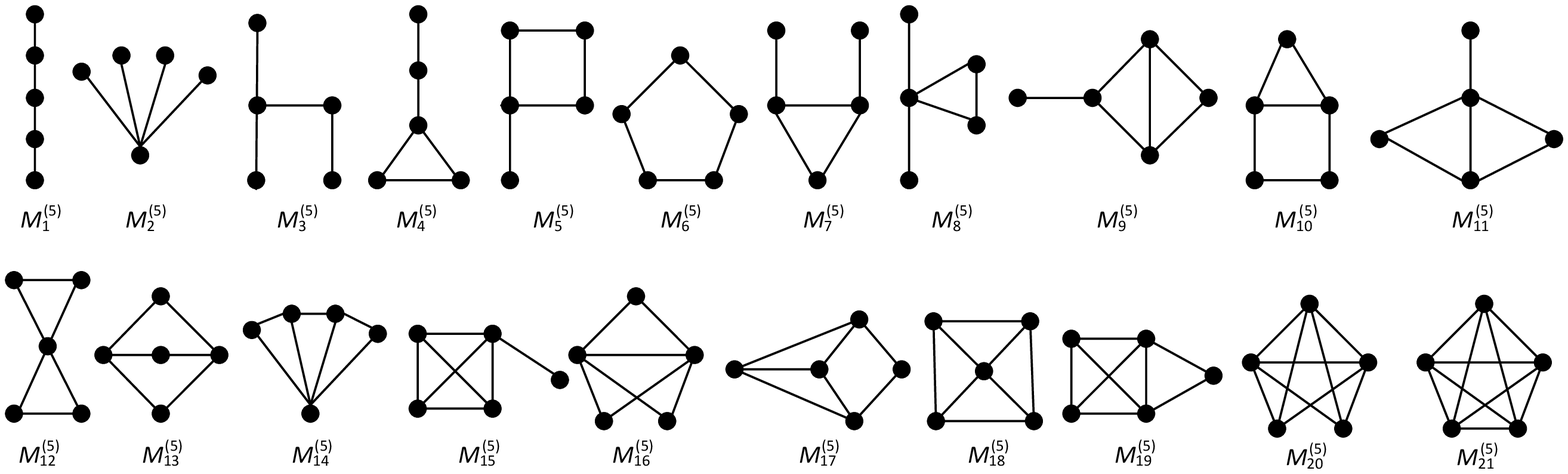}}
\caption{The 4- and 5-node motifs studied in this paper.}
\label{fig:45nodeclasses}
\end{figure*}
Design tools for counting the frequencies of the appearance of 3-, 4-, and 5-node connected subgraph patterns (i.e., motifs, also known as graphlets) in a graph is important for understanding and exploring networks such as online social networks and computer networks.
For example, a variety of motif-based network analysis techniques have been widely used to characterize communication and evolution patterns
in OSNs~\cite{ChunIMC2008,Kunegis2009,ZhaoNetsci2011,Ugander2013},
Internet traffic classification and anomaly detection~\cite{JinSigmetric2009,IliofotouCoNEXT2009},
pattern recognition in gene expression profiling~\cite{Shenorr2002},
protein-protein interaction predication~\cite{Albert2004},
and coarse-grained topology generation~\cite{Itzkovitz2005}.

Due to combinatorial explosion, it is computational intensive to enumerate and count motif frequencies even for a moderately sized graph.
For example, medium-size networks Slashdot~\cite{LeskovecIM2009} and Epinions~\cite{Richardson2003} have $10^5$ nodes and $10^6$ edges but have more than $10^{10}$ 4-node connected and induced subgraphs (CISes)~\cite{TKDDWang2014}.
To address this problem, cheaper methods such as sampling can be used rather than the brute-force enumeration method.
Unfortunately, existing methods of estimating motif concentrations~\cite{Kashtan2004,Wernicke2006,Bhuiyan2012,TKDDWang2014,PinghuiMotifEdgeSampling2014}
cannot be used to estimate motif frequencies, which are more fundamental than motif concentrations.

Despite recent efforts to count triangles~\cite{TsourakakisKDD2009,PavanyVLDB2013,JhaKDD2013,AhmedKDD2014},
little attention has been given to developing scalable tools that can be used to characterize 4- and 5-node motifs.
Jha et al.~\cite{JhaWWW15} develop sampling methods to estimate 4-node undirected motifs' frequencies.
In our experiment we observe that their methods do not bound the estimation error tightly,
so they significantly over-estimate the sampling budget required to achieve a certain accuracy.
Meanwhile, their methods cannot be easily extended to characterize 5-node undirected motifs.
Moreover, their methods use an edge-centric program model, so it is difficult to implement them on current graph computing systems such as Pregel~\cite{Pregel2010}, GraphLab~\cite{LowPVLDB2012} and GraphChi~\cite{KyrolaOSDI12}.
In this paper, we propose new methods to estimate the frequencies of 4- and 5-node motifs.
Our contributions are summarized as:
1) Our methods of sampling 4- and 5-node motifs are computational efficient and scalable.
Meanwile, they can be easily implemented via vertex centric programming models, which are required by most current graph computing systems.
2) To validate our methods, we perform an in-depth analysis.
We find that our methods provide unbiased estimators of motif frequencies.
To the best of our knowledge, we are the first to derive simple and exact formulas for the variances of the estimators,
which is critical for determining a proper sampling budget in practice.
Moreover, we conduct experiments on a variety of publicly available datasets,
and experimental results show that our methods significantly outperform the state-of-the-art methods.

The rest of this paper is organized as follows.
The problem formulation is presented in Section~\ref{sec:problem}.
Section~\ref{sec:preliminaries} introduces preliminaries used in this paper.
Section~\ref{sec:methods} presents our 4- and 5-node motif sampling methods.
The performance evaluation and testing results are presented in Section~\ref{sec:results}.
Section~\ref{sec:related} summarizes related work. Concluding remarks then follow.

\section{Problem Formulation} \label{sec:problem}
Let $G=(V,E)$ be the undirected graph of interest, where $V$ and $E$ are the sets of nodes and edges respectively.
To formally define 4- and 5-node motif frequencies of $G$, we first introduce some notations.
An induced subgraph of $G$, $G'=(V', E')$,
is a subgraph whose edges are \emph{\textbf{all}} in $G$, i.e. $V'\subset V$, $E'=\{(u, v): u, v \in V', (u,v)\in E\}$.
\emph{We would like to point out that if we do not say ``induced" in this paper, we mean that a subgraph is not necessarily induced}.
Fig.~\ref{fig:45nodeclasses}(a) shows all
4-node motifs $M_1^{(4)}, \ldots, M_6^{(4)}$ of any undirected network.
Denote $C_i^{(4)}$ as the set of 4-node CISes in $G$ isomorphic to motif $M_i^{(4)}$,
and then the motif frequency of $M_i^{(4)}$ is defined as $n_i = |C_i^{(4)}|$, $1\le i\le 6$.
Fig.~\ref{fig:45nodeclasses}(b) shows all
5-node motifs $M_1^{(5)}, \ldots, M_{21}^{(5)}$ of any undirected network.
Denote $C_i^{(5)}$ as the set of 5-node CISes in $G$ isomorphic to motif $M_i^{(5)}$,
and then the motif frequency of $M_i^{(5)}$ is defined as $\eta_i = |C_i^{(5)}|$, $1\le i\le 21$.
In this paper, we aim to develop computational methods to estimate $n_1, \ldots, n_6$ and $\eta_1, \ldots, \eta_{21}$.
For ease of reading, we list notations used throughout the paper in Table~\ref{tab:notations}
and we present the proofs of all theorems in this paper in Appendix.

\begin{table}[htb]
\begin{center}
\caption{Table of notations.\label{tab:notations}}
\begin{tabular}{|c|l|} \hline
$G=(V, E)$&$G$ is the undirected graph of interest.\\ \hline
$N_v$&the set of neighbors of a node $v$ in $G$ \\ \hline
$d_v$& $d_v = |N_v|$\\ \hline
$M_1^{(4)}, \ldots, M_6^{(4)}$&4-node undirected motifs\\ \hline
$M^{(4)}(s)$&4-node motif class ID of CIS $s$\\ \hline
\multirow{2}{*}{$C_1^{(4)}, \ldots, C_6^{(4)}$}& $C_i^{(4)}$ as the set of 4-node CISes in $G$\\
&isomorphic to motif $M_i^{(4)}$, $1\le i\le 6$.\\ \hline
\multirow{2}{*}{$n_1^{(4)}, \ldots, n_6^{(4)}$}&the frequency of motif $M_i^{(4)}$, i.e.,\\
&$n_i = |C_i^{(4)}|$, $1\le i\le 6$.\\ \hline
$M_1^{(5)}, \ldots, M_{21}^{(5)}$&5-node undirected motifs\\ \hline
$M^{(5)}(s)$&5-node motif class ID of CIS $s$\\ \hline
\multirow{2}{*}{$C_1^{(5)}, \ldots, C_{21}^{(5)}$}& $C_i^{(5)}$ as the set of 5-node CISes in $G$\\
&isomorphic to motif $M_i^{(5)}$, $1\le i\le 21$.\\ \hline
\multirow{2}{*}{$\eta_1^{(5)}, \ldots, \eta_{21}^{(5)}$}&the frequency of motif $M_i^{(5)}$, i.e.,\\
&$\eta_i = |C_i^{(5)}|$, $1\le i\le 21$.\\ \hline
$N_{u,v}$&$N_{u,v} = \{x: x\in N_u, \text{ and } x\succ v\}$\\ \hline
$d_{u,v}$& $d_{u,v}=|N_{u,v}|$\\ \hline
$K$& sampling budget of MOSS-4\\\hline
$\check K$& sampling budget of MOSS-4Min\\\hline
$K_1$, $K_2$& sampling budgets of MOSS-5\\\hline
\multirow{2}{*}{$\phi_i^{(1)}, 1\le i\le 21$} &the number of subgraphs in motif $M_i^{(5)}$\\
&that are isomorphic to motif $M_3^{(5)}$ \\ \hline
\multirow{2}{*}{$\phi_i^{(2)}, 1\le i\le 21$} &the number of subgraphs in motif $M_i^{(5)}$\\
&that are isomorphic to motif $M_1^{(5)}$ \\ \hline
\multirow{2}{*}{$\phi_i^{(3)}, 1\le i\le 21$} &the number of subgraphs in motif $M_i^{(5)}$\\
&that are isomorphic to motif $M_2^{(5)}$ \\ \hline
$\Omega_1$&$\Omega_1 = \{j: \phi_j^{(1)}>0\}$ \\ \hline
$\Omega_2$&$\Omega_2 = \{j: \phi_j^{(2)}>0\}$ \\ \hline
$\Omega_3$&$\Omega_3 = \{j: \phi_j^{(3)}>0\}$ \\ \hline
\multicolumn{2}{|c|}{$\Lambda_3 =  \sum_{v\in V} \binom{d_v}{3}, \quad \Lambda_4 =  \sum_{v\in V} \binom{d_v}{4}$}\\\hline
\multicolumn{2}{|c|}{$\Gamma = \sum_{v\in V} \left((d_v-1) \sum_{x\in N_v}(d_x - 1)\right)$} \\ \hline
\multicolumn{2}{|c|}{$\check\Gamma = \sum_{v\in V} \sum_{x\in  N_v} d_{v,x} d_{x,v}$} \\ \hline
\multicolumn{2}{|c|}{$\Gamma^{(1)} = \sum_{v\in V} \left((d_v-1)(d_v-2) \sum_{x\in N_v}(d_x - 1)\right)$} \\ \hline
\multicolumn{2}{|c|}{$\Gamma^{(2)} = \sum_{v\in V} \left( \left(\sum_{x\in N_v}(d_x - 1)\right)^2 - \sum_{x\in N_v}(d_x - 1)^2\right)$} \\ \hline
\end{tabular}
\end{center}
\end{table}

\section{Preliminaries}\label{sec:preliminaries}
\subsection{Mix Unbiased Estimators}
\begin{theorem}\label{theorem:mixestimators}
Suppose we have $n$ unbiased estimates $c_1, \ldots, c_n$ of $c$,
i.e., $\mathbb{E} (\hat c_i) = c, i=1,\ldots,n$.
When these estimates are independent and their variances are $\text{Var}(c_i)$, $1\le i\le n$. Using all these estimates, we can obtain a more accurate unbiased estimate $\hat c$ of $c$ by solving
\[
\min_{\sum_{i=1}^n \alpha_i = 1} \text{Var}(\hat c) = \text{Var}(\sum_{i=1}^n \alpha_i c_i).
\]
We can easily obtain the optimal $\text{Var}(\hat c) = \frac{1}{\sum_{j=1}^n \text{Var}^{-1}(c_j)}$ when $\alpha_i = \frac{\text{Var}^{-1}(c_i)}{\sum_{j=1}^n \text{Var}^{-1}(c_j)}$.
We can also estimate the confidence interval of $\hat c$ by the Central Limit Theorem.
That is, as $n\rightarrow +\infty$, for any $\beta>0$, we have
\[
\text{Pr}\left(|\hat c - c| \ge \varepsilon \sqrt{\text{Var}(\hat c)}\right) \rightarrow \frac{1}{\sqrt{2\pi}}\int_{\varepsilon}^{+\infty} e^{-\frac{t^2}{2}} dt \approx \frac{e^{-\frac{\varepsilon^2}{2}}}{\sqrt{2\pi}\varepsilon}.
\]
\end{theorem}

\subsection{3-Path Sampling Methods}
To describe the state-of-the-art 4-node motif sampling methods: 3-path sampling and centered 3-path sampling~\cite{JhaWWW15}, we first introduce some notations.
Let $N_v$ be the set of neighbors of a node $v\in V$ in $G$.
Denote the degree of $v$ as $d_v$,
which is defined as the number of neighbors of $v$ in $G$, i.e., $d_v = |N_v|$.
Let $\succ$ be a total order on all of the nodes in $V$, which can be easily defined and
obtained.
For example, suppose we order all nodes based on their degrees and node IDs, and we define $u\succ v$ if $d_u > d_v$ or, if $d_u = d_v$ while the node ID of $u$ is large than that of $v$.
Let $N_{u,v}$ denote the set of $u$'s neighbors with order larger than $v$, i.e.,
\[
N_{u,v} = \{x: x\in N_u, \text{ and } x\succ v\}.
\]
Denote $d_{u,v}=|N_{u,v}|$.

To sample a 4-node CIS,
the 3-PATH sampling method mainly consists of five steps:
1) Sample an edge $e=(u,v)$ from $E$ according to the distribution
\[
\{\pi_{(u,v)} = \frac{(d_u-1)(d_v-1)}{\sum_{(u',v')\in E} (d_{u'}-1)(d_{v'}-1)}: (u,v)\in E\},
\]
i.e., the probability of sampling an edge $(u,v)\in E$ is $\pi_{(u,v)}$;
2) Sample a node $w$ from $N_v-\{u\}$ uniformly at random;
3) Sample a node $r$ from $N_u-\{v\}$ uniformly at random;
4) Retrieve the CIS $s$ including nodes $v$, $u$, $w$, and $r$.
Note that $s$ might be a 3-node CIS when $r=w$.

Compared to 3-path sampling,
centered 3-path sampling is tailored to estimate the frequencies of 4-node motifs $M_3^{(4)}$, $M_5^{(4)}$, and $M_6^{(4)}$,
which are usually not frequently appeared in many real networks.
To sample a 4-node CIS,
the centered 3-PATH sampling method mainly consists of five steps:
1) Sample an edge $e=(u,v)$ from $E$ according to the distribution
\[
\{\pi_{(u,v)} = \frac{d_{u,v} d_{v,u}}{\sum_{(u',v')\in E} d_{u',v'} d_{v',u'}}: (u',v')\in E\};
\]
2) Sample a node $w$ from $N_{v,u}$ at random;
3) Sample a node $r$ from $N_{u,v}$ at random;
4) Retrieve the CIS $s$ including nodes $v$, $u$, $w$, and $r$.
Similarly, $s$ might be a 3-node CIS.

\subsection{Vertex-Centric Programming Model}
Vertex-centric programming models require users to express their algorithms
by ``thinking like a vertex".
Each node contains information about itself and all its immediate neighbors,
and the algorithms' operations are expressed at the level of a single node.
For example, the operations of a node in Pregel involve receiving messages from other
nodes, updating the state of itself and its edges, and sending messages
to other nodes.
Vertex-centric models are very easy to program and have been
widely used for many graph mining and machine learning algorithms.

\section{Sampling 4-Node Motifs} \label{sec:methods}
In this section, we introduce our sampling methods: MOSS-4 and MOSS-4Min.
MOSS-4 is used to estimate all 4-node motifs' frequencies.
We observe that MOSS-4 might exhibit large errors for characterizing rare motifs (i.e., motifs with low frequencies) for a small sampling budget.
In addition to MOSS-4, we also develop a method MOSS-4Min to further reduce the errors for characterizing rare motifs.

\subsection{MOSS-4}
\subsubsection{Sampling}
Denote by $\Gamma_v = (d_v-1) \sum_{x\in N_v}(d_x - 1)$.
We assign a weight $\Gamma_v$ to each node $v\in V$.
Define $\Gamma = \sum_{v\in V} \Gamma_v$ and $\pi_v = \frac{\Gamma_v}{\Gamma}$.
Our method of sampling a 4-node CIS mainly consists of five steps:
1) Sample a node $v$ from $V$ according to the distribution $\pi = \{\pi_v: v\in V\}$;
2) Sample a random node $u$ from $N_v$ according to the distribution $\sigma^{(v)}=\{\sigma_u^{(v)}: u\in N_v\}$,
where $\sigma_u^{(v)}$ is defined as
\begin{equation}\label{eq:sigmaNv}
\sigma_u^{(v)} = \frac{d_u-1}{\sum_{x\in N_v} (d_x-1)}, \quad u\in N_v;
\end{equation}
3) Sample a node $w$ from $N_v-\{u\}$ uniformly at random;
4) Sample a node $r$ from $N_u-\{v\}$ uniformly at random;
5) Retrieve the CIS $s$ including nodes $v$, $u$, $w$, and $r$.
We set the sampling budget as $K$, i.e., we run the above method $K$ times to obtain $K$ CISes $s_1, \ldots, s_K$.
The pseudo-code of MOSS-4 is shown in Algorithm~\ref{alg:MOSS4}.
In Algorithm~\ref{alg:MOSS4}, function $\text{WeightRandomVertex}(V, \pi)$ returns a node sampled from $V$ according to the distribution $\pi=\{\pi_v: v\in V\}$,
function $\text{RandomVertex}(X)$ returns a node sampled from $X$ at random,
and function $\text{CIS}(\{v,u,w,r\})$ returns the CIS with the node set $\{v,u,w,r\}$ in $G$.

\subsubsection{Estimator}
Let $\varphi_i^{(1)}$, $1\le i\le 6$, be the number of subgraphs in motif $M_i^{(4)}$ that are isomorphic to motif $M_1^{(4)}$.
We can easily compute $\varphi_1^{(1)} = 1$, $\varphi_2^{(1)} = 0$, $\varphi_3^{(1)} = 4$, $\varphi_4^{(1)} = 2$, $\varphi_5^{(1)} = 6$, and $\varphi_6^{(1)} = 12$.
To remove the error introduced by sampling,
we analyze the bias of MOSS-4 as follows:
\begin{theorem}\label{theorem: MOSS4samplingweight}
When the sampling budget $K=1$,
MOSS-4 samples a CIS $s\in C_i^{(4)}$ with probability $p_i = \frac{2\varphi_i^{(1)}}{\Gamma}$, $1\le i\le 6$.
\end{theorem}

\begin{algorithm}
\SetKwRepeat{Do}{do}{while}%
\SetKwFunction{CIS}{CIS}
\SetKwFunction{WeightRandomVertex}{WeightRandomVertex}
\SetKwFunction{RandomVertex}{RandomVertex}
\SetKwInOut{Input}{input}
\SetKwInOut{Output}{output}
\tcc{\footnotesize $K$ is the sampling budget.}
\Input{$G=(V, E)$ and $K$.}
\Output{$\hat n_i$, $1\le i\le 6$.}
\BlankLine

\For {$i \in \{1,3,4,5,6\}$}{
$\hat n_i \gets 0$\;
}

\For {$k \in [1, K]$} {
    $v \gets \WeightRandomVertex(V, \pi)$\;
    $u \gets \WeightRandomVertex(N_v, \sigma^{(v)})$\;
    $w \gets \RandomVertex(N_v-\{u\})$\;
    $r \gets \RandomVertex(N_u-\{v\})$\;
    $s_k\gets  \CIS(\{v,u,w,r\})$\;
    \If {$r \ne u$ and $r \ne w$} {
        $i \gets M^{(4)}(s_k)$\;
        $\hat n_i \gets \hat n_i + \frac{1}{K p_i}$\;
    }
}

$\hat n_2 \gets \Lambda_3 - \hat n_4 - 2 \hat n_5 - 4 \hat n_6$\;
\caption{The pseudo-code of MOSS-4. \label{alg:MOSS4}}
\end{algorithm}

We let $M^{(4)}(s_k)$ be the 4-node motif class ID of $s_k$ when $s_k$ is a 4-node CIS, and -1 otherwise (i.e., $s_k$ is a triangle).
Let $\mathbf{1}(\mathbb{X})$ denote the indicator function that equals one when predicate $\mathbb{X}$ is true, and zero otherwise.
Denote $m_i=\sum_{k=1}^{K} \mathbf{1}(M^{(4)}(s_k)=i)$.
For $i\in \{1, 3, 4, 5, 6\}$, $p_i$ is larger than zero and we estimate $n_i$ as
\[
\hat n_i = \frac{m_i}{K p_i}, \quad i\in \{1, 3, 4, 5, 6\}.
\]
Let $\Lambda_3 =  \sum_{v\in V} \binom{d_v}{3}$.
Then, the number of all 4-node subgraphs (not necessarily induced) in $G$ isomorphic to motif $M_2^{(4)}$ is $\Lambda_3$.
Let $\varphi_i^{(2)}$, $1\le i\le 6$, be the number of subgraphs in motif $M_i^{(4)}$ that are isomorphic to motif $M_2^{(4)}$.
We have $\varphi_1^{(2)} = 0$, $\varphi_2^{(2)} = 1$, $\varphi_3^{(2)} = 0$, $\varphi_4^{(2)} = 1$, $\varphi_5^{(2)} = 2$, and $\varphi_6^{(2)} = 4$.
We can easily find that
\begin{equation}\label{eq:n2}
\Lambda_3  = \sum_{i=1}^6 \varphi_i^{(2)} n_i = n_2 + n_4 + 2 n_5 + 4 n_6.
\end{equation}
Thus, we estimate $n_2$ as

\[
\hat n_2 = \Lambda_3 - \hat n_4 - 2 \hat n_5 - 4 \hat n_6.
\]

\begin{theorem}\label{theorem: varianceMOSS4}
$\hat n_i$ is an unbiased estimator of $n_i$, $1\le i\le 6$.
The variance of $\hat n_i$ is
\[
\text{Var} (\hat n_i) = \frac{n_i}{K} \left(\frac{1}{p_i} - n_i\right), \quad i\in \{1, 3, 4, 5, 6\}.
\]
The variance of $\hat n_2$ is computed as
\[
\text{Var} (\hat n_2) = \frac{1}{K} \left(\frac{n_4}{p_4} + \frac{4 n_5}{p_5} + \frac{16 n_6}{p_6} -(n_4 + 2 n_5 + 4 n_6)^2\right).
\]
\end{theorem}
From Theorem~\ref{theorem:mixestimators},
we can easily compute a sampling budget $K$ that can guarantee $P(|\hat n_i -n_i|>\varepsilon n_i)<\delta$ for any $\varepsilon > 0$ and $0<\delta<1$, $i=1,\ldots, 6$.

\subsubsection{Computational Complexity} \label{sec:samplingfunctions}
\textbf{Initialization}:
For each node $v$, we store its degree $d_v$ and use a list to store its neighbors $N_v$.
Therefore, it requires $O(d_v)$ operations to compute $\Gamma_v$,
and the computational complexity of processing all nodes is $O(|E|)$.

\textbf{$\text{WeightRandomVertex}(V, \pi)$}:
We use a list $V[1,\ldots, |V|]$ to store the nodes in $V$.
We store an array $ACC\_\Gamma[1,\ldots, |V|]$ in memory, where $ACC\_\Gamma[i]$ is defined as $ACC\_\Gamma[i] = \sum_{j=1}^i \Gamma_{V[j]}$, $1\le i\le |V|$.
Clearly, $ACC\_\Gamma[|V|] = \Gamma$.
Let $ACC\_\Gamma[0]=0$.
Then, $\text{WeightRandomVertex}(V, \pi)$ can be easily achieved by the following three steps:
\begin{itemize}
  \item Step 1: Select a number $rnd$ from $\{1, \ldots, \Gamma\}$ at random;
  \item Step 2: Find $i$ satisfying $ACC\_\Gamma[i-1] < rnd \le ACC\_\Gamma[i]$,
which can be solved by the binary search algorithm;
  \item Step 3: Return $V[i]$.
\end{itemize}
Its computational complexity is $O(\log|V|)$.

\textbf{$\text{WeightRandomVertex}(N_v, \sigma^{(v)})$}:
We use a list $N_v[1, \ldots, d_v]$ to store the neighbors of $v$. 
We store an array $ACC\_\sigma^{(v)}[1, \ldots, d_v]$ in memory,
where $ACC\_\sigma^{(v)}[i]$ is defined as $ACC\_\sigma^{(v)}[i] = \sum_{j=1}^i (d_{N_v[j]} - 1)$, $1\le i\le d_v$.
Let $ACC\_\sigma^{(v)}[0]=0$.
Then, $\text{WeightRandomVertex}(N_v, \sigma^{(v)})$ can be easily achieved by the following three steps:
\begin{itemize}
  \item Step 1: Select a number $rnd$ from $\{1, \ldots, ACC\_\sigma^{(v)}[d_v]\}$ at random;
  \item Step 2: Find $i$ satisfying 
  \[
  ACC\_\sigma^{(v)}[i-1] < rnd \le ACC\_\sigma^{(v)}[i],
  \]
which can be solved by the binary search algorithm;
  \item Step 3: Return $N_v[i]$.
\end{itemize}
Its computational complexity is $O(\log d_v)$. 

\textbf{$\text{RandomVertex}(N_v-\{u\})$}: Let $POS_{v,u}$ denote the index of $u$ in the list $N_v[1, \ldots, d_v]$, i.e., $N_v[POS_{v,u}]=u$.
Then, function $\text{RandomVertex}(N_v-\{u\})$ can be achieved by the following steps:
\begin{itemize}
  \item Step 1: Select a number $rnd$ from $\{1, \ldots, d_v\} - \{POS_{v,u}\}$ at random;
  \item Step 2: Return $N_v[rnd]$.
\end{itemize}
Its computational complexity is $O(1)$.

In summary, the complexity of MOSS-4 sampling $K$ CISes is $O(|E|+K\log|V|)$.

\subsection{MOSS-4Min}
\subsubsection{Sampling}
From the above derived formulas of the variances of MOSS-4,
we can see that MOSS-4 might exhibit larger errors for 4-node motifs with lower frequencies when allocating a small sampling budget $K$.
To solve this problem, we develop a better method MOSS-4Min to further reduce the errors for estimating the frequencies of 4-node motifs $M_3^{(4)}$, $M_5^{(4)}$, and $M_6^{(4)}$.

Let $\check \Gamma_v = \sum_{x\in  N_v} d_{v,x} d_{x,v}$, $v\in V$.
MOSS-4Min assigns a weight $\check \Gamma_v$ to each node $v\in V$.
Define $\check \Gamma = \sum_{v\in V} \check \Gamma_v$ and $\check \pi_v = \frac{\check \Gamma_v}{\check \Gamma}$.
MOSS-4Min mainly consists of five steps:
1) Sample a node $v$ from $V$ according to the distribution $\check \pi = \{\check \pi_v: v\in V\}$.
2) Sample a node $u$ from $N_v$ according to the distribution $\check \sigma^{(v)}=\{\check \sigma_u^{(v)}: u\in N_v\}$,
where $\check \sigma_u^{(v)}$ is defined as
\begin{equation}\label{eq:checksigmaNv}
\check \sigma_u^{(v)} = \frac{d_{u,v} d_{v,u}}{\check \Gamma_v}, \quad u\in N_v;
\end{equation}
3) Sample a node $w$ from $N_{v,u}$ at random;
4) Sample a node $r$ from $N_{u,v}$ at random;
5) Retrieve the CIS $s$ including nodes $v$, $u$, $w$, and $r$.
We set the sampling budget as $\check K$ to obtain $\check K$ CISes $s_1, \ldots, s_{\check K}$.

\subsubsection{Estimator}

\begin{algorithm}
\SetKwRepeat{Do}{do}{while}%
\SetKwFunction{CIS}{CIS}
\SetKwFunction{OrbitMin}{OrbitMin}
\SetKwFunction{WeightRandomVertex}{WeightRandomVertex}
\SetKwFunction{RandomVertex}{RandomVertex}
\SetKwInOut{Input}{input}
\SetKwInOut{Output}{output}
\Input{$G=(V, E)$ and $\check K$.}
\Output{$\check n_i$, $i \in \{3,5,6\}$.}
\BlankLine

$\check n_3 \gets 0$, $\check n_5 \gets 0$, and $\check n_6 \gets 0$\;

\For {$k \in [1, \check K]$} {
    $v \gets \WeightRandomVertex(V, \check\pi)$\;
    $u \gets \WeightRandomVertex(N_v, \check \sigma^{(v)})$\;
    $w \gets \RandomVertex(N_{v,u})$\;
    $r \gets \RandomVertex(N_{u,v})$\;
    $s_k\gets  \CIS(\{v,u,w,r\})$\;
    \If {$r \ne u$ and $r \ne w$} {
        $i \gets M^{(4)}(s_k)$\;
        \If{$i\in \{3, 5, 6\}$} {
            $\check n_i \gets \check n_i + \frac{1}{\check K \check p_i}$\;
        }
    }
}
\caption{The pseudo-code of MOSS-4Min. \label{alg:MOSS4Min}}
\end{algorithm}

\begin{theorem}\label{theorem: MOSS4Minsamplingweight}
When the sampling budget $\check K =1$, MOSS-4Min samples CISes $s\in C_3^{(4)}$, $s\in C_3^{(5)}$, and $s\in C_6^{(4)}$ with probabilities $\check p_3 = 2 \check \Gamma^{-1}$, $\check p_5 = 2 \check \Gamma^{-1}$, and $\check p_6 = 6 \check \Gamma^{-1}$ respectively.
\end{theorem}

We estimate $n_3$, $n_5$, and $n_6$ as
\[
\check n_i = \sum_{k=1}^{\check K} \frac{m_i}{\check K \check p_i}, \quad i=3, 5, 6,
\]
where $m_i=\sum_{k=1}^{K} \mathbf{1}(M^{(4)}(s_k)=i)$.
The variances of $\check n_3$, $\check n_5$, and $\check n_6$ are given in the following theorem.
We omit the proof, which is analogous to that of Theorem~\ref{theorem: varianceMOSS4}.

\begin{theorem}\label{theorem: varianceMOSS4Min}
$\check n_i$ is an unbiased estimator of $n_i$, $i=3,5,6$. Its variance is
\[
\text{Var} (\check n_i) = \frac{n_i}{\check K} \left(\frac{1}{\check p_i} - n_i\right), \quad i=3, 5, 6.
\]
\end{theorem}

From Theorems~\ref{theorem:mixestimators},~\ref{theorem: varianceMOSS4}, and~\ref{theorem: varianceMOSS4Min}, we can easily obtain a more accurate estimator of $n_i$ by combining $\hat n_i$ and $\check n_i$, $i=3, 5, 6$.

\subsubsection{Computational Complexity}
We easily extend methods in Section~\ref{sec:samplingfunctions} to design functions $\text{WeightRandomVertex}(V, \check\pi)$ and $\text{WeightRandomVertex}(N_v, \check \sigma^{(v)})$ in Algorithm~\ref{alg:MOSS4Min}.
The computational complexity of MOSS-4min sampling $K$ CISes is $O(|E|+K\log|V|)$.

\subsection{Vertex-Centric Programming Models}
In this subsection, we show MOSS-4 and MOSS-4MIN can be easily implemented via vertex-centric programming models.
\subsubsection{Vertex-Centric Programming Model of MOSS-4 Sampling Method}
First, we sample $K$ nodes in $V$ according to $\pi$.
Let $k_v$ denote the number of times a node $v\in V$ sampled.
Thus, $\sum_{v\in V} k_v = K$.
For each node $v$, we set $k_v$ as its node value, and then repeat the set of four following operations $k_v$ times
\[
u\leftarrow \text{WeightRandomVertex}(N_v, \sigma^{(v)}),
\]
\[
w\leftarrow  \text{RandomVertex}(N_v-\{u\}),
\]
\[
\text{Update}(A) \text{ and then } \text{MSG}(v,*,w,*,A)\rightarrow u,
\]
where $A$ is the adjacent matrix of the CIS consisting of nodes $v$, $u$, $w$, and $r$,
which are the variables in the Algorithm~\ref{alg:MOSS4}, i.e., the nodes sampled at the 1-st, 2-nd, 3-rd, and 4-th steps respectively.
Note that here $r$ and some entries in $A$ are unknown.
Function $\text{Update}(A)$ is used to get the values of unknown entries in $A$ based the edges of the current node $v$.
Function $\text{MSG}(v,*,w,*,A)\rightarrow u$ generates a message $(v,*,w,*,A)$, and sends the message to $u$, which is a neighbor of $v$.

We process the messages that a node receives as follows:
\begin{itemize}
  \item When a node $u$ receives a message like $(v,*,w,*,A)$, do
\[
r \leftarrow \text{RandomVertex}(N_u-\{v\}),
\]
\[
\text{Update}(A) \text{ and then } \text{MSG}(v,u,w,*,A) \rightarrow r.
\]
  \item When a node $r$ receives a message like $(v,u,w,*,A)$, we first $\text{Update}(A)$.
  From $A$, we then have all the edges between $v$, $u$, $w$, and $r$. Last, we set $m_i\leftarrow m_i+1$, where $i$ is the motif class of the CIS consisting of $v$, $u$, $w$, and $r$.
\end{itemize}

\subsubsection{Vertex-Centric Programming Model of MOSS-4Min Sampling Method}
Similar to MOSS-4, we sample $\check K$ nodes in $V$ according to $\check\pi$.
Let $\check k_v$ denote the number of times a node $v\in V$ sampled.
Thus, $\sum_{v\in V} \check k_v = \check K$.
For each node $v$, we set $\check k_v$ as its node value, and then repeat the set of four following operations $\check k_v$ times
\[
u\leftarrow \text{WeightRandomVertex}(N_v, \check \sigma^{(v)}),
\]
\[
w\leftarrow  \text{RandomVertex}(N_{v, u}),
\]
\[
\text{Update}(A) \text{ and then } \text{MSG}(v,*,w,*,A)\rightarrow u.
\]

We process the messages that a node receives as follows:
\begin{itemize}
  \item When a node $u$ receives a message like $(v,*,w,*,A)$, do
\[
r \leftarrow \text{RandomVertex}(N_{u,v}),
\]
\[
\text{Update}(A) \text{ and then } \text{MSG}(v,u,w,*,A) \rightarrow r.
\]
  \item When a node $r$ receives a message like $(v,u,w,*,A)$, we first $\text{Update}(A)$ and then set $m_i\leftarrow m_i+1$, where $i$ is the motif class of the CIS consisting of $v$, $u$, $w$, and $r$.
\end{itemize}

\subsection{Relationship to 3-Path Sampling and Centered 3-Path Sampling}
MOSS-4 and MOSS4-Min can be viewed as the vertex-centric versions of the 3-path and centered 3-path sampling methods respectively.
Suppose we use 4 bytes to store a node ID and its weight $\Gamma_v$.
The 3-path and centered 3-path sampling methods require $8|E| + 4d_{max}$ bytes of memory, but MOSS-4 and MOSS-4Min need only $4(|V| + d_{max})$ bytes,
which is orders of magnitude smaller than $8|E| + 4d_{max}$ for many real-world large networks.
Therefore, MOSS-4 and MOSS-4Min are suit for disk-based graph computing systems such as GraphChi and VENUS~\cite{ChengLLFLH15},
which aim to analyze big graphs when the graphs of interest cannot be fitted into memory.
Moreover, MOSS-4 and MOSS-4Min can be easily implemented in distributed vertex-centric graph computing systems such as Pregel and GraphLab.
Meanwhile, we would like to point out we give the closed-form formulas for the variances of MOSS-4 and MOSS-4Min.
They are critical to evaluate the error of an estimate and determine a proper sampling budget in order to guarantee certain accuracy.
Moreover, they can also help us to make the right sampling strategies in advance.
An example is given in the following subsection.

\subsection{Compare MOSS-4 and MOSS-4Min}\label{subsection: compareMOSS4}
From Theorems~\ref{theorem: varianceMOSS4} and~\ref{theorem: varianceMOSS4Min},
when $K=\check K$, we have
\[
\frac{\text{Var} (\hat n_i)} {\text{Var} (\check n_i)} = \frac{1/p_i - n_i} {1/\check p_i - n_i} \approx \frac{\check p_i} {p_i}, \quad, i=3, 5, 6,
\]
where $\frac{\check p_3} {p_3} = \frac{\Gamma}{4 \check \Gamma}$, $\frac{\check p_5}{p_5}=\frac{\Gamma} {6 \check \Gamma}$,
and $\frac{\check p_6}{p_6}  = \frac{\Gamma} {4 \check \Gamma}$.
Thus, the value of $\frac{\Gamma} {\check\Gamma}$ helps us to determine
whether it is necessary to apply MOSS-4Min to further reduce the errors of estimating $n_3$, $n_5$, and $n_6$.
For example, the graph ca-GrQc~\cite{LeskovecTKDD2007} has $\frac{\Gamma} {\check\Gamma}=5.5$.
In our experiments we observe that MOSS-4Min slightly improves the accuracy of MOSS-4 for estimating $n_3$ and $n_6$ of ca-GrQc,
and exhibits a larger error than MOSS-4 for estimating $n_5$ of ca-GrQc.

\section{Sampling 5-Node Motifs}\label{sec:MOSS-5}
\subsection{MOSS-5}
MOSS-5, our method of estimating frequency of all 5-node motifs, consists of two sub-methods: T-5 and Path-5.
We develop T-5 to sample 5-node CISes that include at least one subgraph isomorphic to $M_3^{(5)}$.
Similarly, Path-5 is developed to sample 5-node CISes that include at least one subgraph isomorphic to $M_1^{(5)}$.
Finally, we propose a method to estimate the frequency of all 5-node motifs based on sampled CISes given by T-5 and Path-5.
\subsubsection{T-5 Sampling Method}
The pseudo-code of T-5 is shown in Algorithm~\ref{alg:T-5}.
Let
\[
\Gamma_v^{(1)} = (d_v-1)(d_v-2) \sum_{x\in N_v}(d_x - 1), \quad v\in V.
\]
We assign a weight $\Gamma_v^{(1)}$ to each node $v\in V$.
Define $\Gamma^{(1)} = \sum_{v\in V} \Gamma_v^{(1)}$ and $\rho_v^{(1)} = \frac{\Gamma_v^{(1)}}{\Gamma^{(1)}}$.
To sample a 5-node CIS,
T-5 mainly consists of five steps:
1) Sample a node $v$ from $V$ according to the distribution $\rho^{(1)} = \{\rho_v^{(1)}: v\in V\}$;
2) Sample a node $u$ of $N_v$ according to the distribution $\sigma^{(v)}=\{\sigma_u^{(v)}: u\in N_v\}$,
where $\sigma_u^{(v)}$ is defined the same as in~(\ref{eq:sigmaNv});
3) Sample two different nodes $w$ and $r$ from $N_v-\{u\}$ at random;
4) Sample a node $t$ from $N_u-\{v\}$ uniformly at random;
5) Retrieve the CIS $s$ including nodes $v$, $u$, $w$, $r$ and $t$.
We run the above method $K_1$ times to obtain $K_1$ CISes $s_1^{(1)}, \ldots, s_{K_1}^{(1)}$.

\begin{algorithm}
\SetKwRepeat{Do}{do}{while}%
\SetKwFunction{CIS}{CIS}
\SetKwFunction{WeightRandomVertex}{WeightRandomVertex}
\SetKwFunction{RandomVertex}{RandomVertex}
\SetKwInOut{Input}{input}
\SetKwInOut{Output}{output}
\Input{$G=(V, E)$ and $K_1$.}
\Output{$\hat \eta_i^{(1)}$.}
\BlankLine

\For {$i \in \Omega_1$}{
    $\hat \eta_i^{(1)} \gets 0$\;
}

\For {$k \in [1, K_1]$}{
    $v \gets \WeightRandomVertex(V, \rho^{(1)})$\;
    $u \gets \WeightRandomVertex(N_v, \sigma^{(v)})$\;
    $w \gets \RandomVertex(N_v - \{u\})$\;
    $r \gets \RandomVertex(N_v -\{u, w\})$\;
    $t \gets \RandomVertex(N_u - \{v\})$\;
    $s_k^{(1)}\gets  \CIS(\{v,u,w,r,t\})$\;
    \If {$t \ne w$ and $t \ne r$} {
        $i \gets M^{(5)}(s_k^{(1)})$\;
        $\hat \eta_i^{(1)} \gets \hat \eta_i^{(1)} + \frac{1}{K_1 p_i^{(1)}}$\;
    }
}
\caption{The pseudo-code of T-5. \label{alg:T-5}}
\end{algorithm}

Let $\phi_i^{(1)}$, $1\le i\le 21$, be the number of subgraphs in motif $M_i^{(5)}$ that are isomorphic to motif $M_3^{(5)}$.
The value of $\phi_i^{(1)}$ is given in Table~\ref{tab:5nodephi}.
The following theorem shows the sampling bias of the 5-node T-sampling method.
\begin{theorem}\label{theorem: Tsamplersamplingweight}
When the sampling budget $K_1=1$,
T-5 samples a CIS $s\in C_i^{(5)}$ with probability $p_i^{(1)} = \frac{2\phi_i^{(1)}}{\Gamma^{(1)}}$, $1\le i\le 21$.
\end{theorem}

\begin{table*}[htb]
\begin{center}
\caption{Values of $\phi_{i}^{(1)}$, $\phi_{i}^{(2)}$, and $\phi_{i}^{(3)}$.\label{tab:5nodephi}}
\begin{tabular}{|r|rrrrrrrrrrrrrrrrrrrrr|} \hline
$i$&1&2&3&4&5&6&7&8&9&10&11&12&13&14&15&16&17&18&19&20&21\\
\hline
$\phi_{i}^{(1)}$&0&0&1&1&2&0&2&2&4&4&5&4&6&10&9&12&10&20&20&36&60\\
$\phi_{i}^{(2)}$&1&0&0&2&2&5&1&0&4&7&2&4&6&10&6&6&14&24&18&36&60\\
$\phi_{i}^{(3)}$&0&1&0&0&0&0&0&1&0&0&1&1&0&1&1&2&0&1&2&3&5\\
\hline
\end{tabular}
\end{center}
\end{table*}

We let $M^{(5)}(s)$ be the 5-node motif class ID of $s$ when $s$ is a 5-node CIS, and -1 otherwise.
Denote $m_i^{(1)} = \sum_{k=1}^{K_1} \mathbf{1}(M^{(5)}(s_k^{(1)})=i)$.
Let $\Omega_1 = \{j: \phi_j^{(1)}>0\}$.
For $i\in \Omega_1$, $p_i^{(1)}$ is larger than zero and we then estimate $\eta_i$ as
\[
\hat \eta_i^{(1)} = \frac{m_i^{(1)}}{K_1 p_i^{(1)}}.
\]

\begin{theorem}\label{theorem: varianceT-5}
For $i\in \Omega_1$, $\hat \eta_i^{(1)}$ is an unbiased estimator of $\eta_i$
and its variance of $\hat \eta_i^{(1)}$ is
\begin{equation}\label{eq:variance_T-5}
\text{Var} (\hat \eta_i^{(1)}) = \frac{\eta_i}{K_1} \left(\frac{1}{p_i^{(1)}} - \eta_i\right).
\end{equation}
The covariance of $\hat \eta_i^{(1)}$ and $\hat \eta_j^{(1)}$  is
\[
\text{Cov}(\hat \eta_i^{(1)}, \hat \eta_j^{(1)}) = -\frac{\eta_i \eta_j}{K_1}, \quad i\ne j \text{ and } i, j\in \Omega_1.
\]
\end{theorem}

\subsubsection{Path-5 Sampling Method}
The pseudo-code of Path-5 is shown in Algorithm~\ref{alg:Path-5}.
Let
\[
\Gamma_v^{(2)} = \left(\sum_{x\in N_v}(d_x - 1)\right)^2 - \sum_{x\in N_v}(d_x - 1)^2, \quad v\in V.
\]
We assign a weight $\Gamma_v^{(2)}$ to each node $v\in V$.
Define $\Gamma^{(2)} = \sum_{v\in V} \Gamma_v^{(2)}$ and $\rho_v^{(2)} = \frac{\Gamma_v^{(2)}}{\Gamma^{(2)}}$.
To sample a 5-node CIS,
Path-5 mainly consists of six steps:
1) Sample a node $v$ from $V$ according to the distribution $\rho^{(2)} = \{\rho_v^{(2)}: v\in V\}$;
2) Sample a node $u$ from $N_v$ according to the distribution $\tau^{(v)}=\{\tau_u^{(v)}: u\in N_v\}$,
where $\sum_{u\in N_v}\tau_u^{(v)} = 1$ and $\tau_u^{(v)}$ is defined as
\[
\tau_u^{(v)} = \frac{(d_u - 1) (\sum_{y\in N_v-\{u\}} (d_y - 1))}{\Gamma_v^{(2)}}, \quad u\in N_v;
\]
3) Sample a node $w$ from $N_v-\{u\}$ according to the distribution $\mu^{(v, u)}=\left\{\mu_w^{(v, u)}: w\in N_v-\{u\}\right\}$,
where $\sum_{w\in N_v-\{u\}} \mu_w^{(v, u)} = 1$ and $\mu_w^{(v, u)}$ is defined as
\[
\mu_w^{(v, u)} = \frac{d_w - 1}{\sum_{y\in N_v-\{u\}} (d_y - 1)}, \quad w\in N_v-\{u\};
\]
4) Sample a node $r$ from $N_u-\{v\}$ uniformly at random;
5) Sample a node $t$ from $N_w-\{v\}$ uniformly at random;
6) Retrieve the CIS $s$ including nodes $v$, $u$, $w$, $r$ and $t$.
We run the above method $K_2$ times to obtain $K_2$ CISes $s_1^{(2)}, \ldots, s_{K_2}^{(2)}$.

\begin{algorithm}
\SetKwRepeat{Do}{do}{while}%
\SetKwFunction{CIS}{CIS}
\SetKwFunction{WeightRandomVertex}{WeightRandomVertex}
\SetKwFunction{RandomVertex}{RandomVertex}
\SetKwInOut{Input}{input}
\SetKwInOut{Output}{output}
\Input{$G=(V, E)$ and $K_2$.}
\Output{$\hat \eta_i^{(2)}$.}
\BlankLine

\For {$i \in \Omega_2$}{
    $\hat \eta_i^{(2)} \gets 0$\;
}
\For {$k \in [1, K_2]$} {
    $v \gets \WeightRandomVertex(V, \rho^{(2)})$\;
    $u \gets \WeightRandomVertex(N_v, \tau^{(v)})$\;
    $w \gets \WeightRandomVertex(N_v- \{u\}, \mu^{(v, u)})$\;
    $r \gets \RandomVertex(N_u-\{v\})$\;
    $t \gets \RandomVertex(N_w-\{v\})$\;
    $s_k^{(2)}\gets  \CIS(\{v,u,w,r,t\})$\;
    \If {$t \ne u$ and $r \ne w$ and $t \ne r$} {
        $i \gets M^{(5)}(s_k^{(2)})$\;
        $\hat \eta_i^{(2)} \gets \hat \eta_i^{(2)} + \frac{1}{K_2 p_i^{(2)}}$\;
    }

}
\caption{The pseudo-code of Path-5. \label{alg:Path-5}}
\end{algorithm}

Let $\phi_i^{(2)}$, $1\le i\le 21$, be the number of subgraphs in motif $M_i^{(5)}$ that are isomorphic to motif $M_1^{(5)}$.
The value of $\phi_i^{(2)}$ is given in Table~\ref{tab:5nodephi}.
The following theorem shows the sampling bias of Path-5.

\begin{theorem}\label{theorem: Path-5samplingweight}
When the sampling budget $K_2=1$,
Path-5 samples a CIS $s\in C_i^{(5)}$ with probability $p_i^{(2)} = \frac{2\phi_i^{(2)}}{\Gamma^{(2)}}$, $1\le i\le 21$.
\end{theorem}

Denote $m_i^{(2)} = \sum_{k=1}^{K_2} \mathbf{1}(M^{(5)}(s_k^{(2)})=i)$.
Let $\Omega_2 = \{j: \phi_j^{(2)}>0\}$.
For $i\in \Omega_2$, $p_i^{(2)}$ is larger than zero and we then estimate $\eta_i$ as
\[
\hat \eta_i^{(2)} = \frac{m_i^{(2)}}{K_2 p_i^{(2)}}.
\]

\begin{theorem}\label{theorem: variancePath-5}
For $i\in \Omega_2$, $\hat \eta_i^{(2)}$ is an unbiased estimator of $\eta_i$
and its variance of $\hat \eta_i^{(2)}$ is
\begin{equation}\label{eq:variance_Path-5}
\text{Var} (\hat \eta_i^{(2)}) = \frac{\eta_i}{K_2} \left(\frac{1}{p_i^{(2)}} - \eta_i\right).
\end{equation}
The covariance of $\hat \eta_i^{(2)}$ and $\hat \eta_j^{(2)}$  is
\[
\text{Cov}(\hat \eta_i^{(2)}, \hat \eta_j^{(2)}) = -\frac{\eta_i \eta_j}{K_2}, \quad i\ne j \text{ and } i, j\in \Omega_2.
\]
\end{theorem}

\subsubsection{Mix Estimator}
We estimate $\eta_i$ as $\hat \eta_i^{(1)}$ and $\hat \eta_i^{(2)}$ for $i\in \Omega_1- \Omega_2$ and $i\in \Omega_2- \Omega_1$ respectively.
When $i\in \Omega_1\cap \Omega_2$, according to Thereom~\ref{theorem:mixestimators}, we estimate $\eta_i$ based on its two estimates $\hat \eta_i^{(1)}$ and $\hat \eta_i^{(2)}$.
Formally, we define
\[
\lambda_i^{(1)} =\frac{\text{Var}(\hat \eta_i^{(2)})}{\text{Var}(\hat \eta_i^{(1)}) + \text{Var}(\hat \eta_i^{(2)})}\text{ and } \lambda_i^{(2)} = \frac{\text{Var}(\hat \eta_i^{(1)})}{\text{Var}(\hat \eta_i^{(1)}) + \text{Var}(\hat \eta_i^{(2)})}
\]
where $\text{Var}(\hat \eta_i^{(1)})$ and $\text{Var}(\hat \eta_i^{(2)})$ are given in~(\ref{eq:variance_T-5}) and~(\ref{eq:variance_Path-5}).
For $i\in \Omega_1 \cup \Omega_2 = \{1,3,4,5,\ldots,21\}$, we finally estimate $\eta_i$ as
\begin{equation}\label{eq:mixestimator}
\hat \eta_i = \begin{dcases}
\lambda_i^{(1)} \hat \eta_i^{(1)}  + \lambda_i^{(2)} \hat \eta_i^{(2)}, & i\in \Omega_1 \cap \Omega_2, \\
\hat \eta_i^{(1)}, & i\in \Omega_1 - \Omega_2,\\
\hat \eta_i^{(2)}, & i\in \Omega_2 - \Omega_1.
\end{dcases}
\end{equation}

We can see that $\Omega_1\cup \Omega_2 = \{1,2,\ldots,21\} - \{2\}$.
Thus,~(\ref{eq:mixestimator}) can be used to estimate the frequencies of all 5-node motifs except motif $M_2^{(5)}$.
Next, we introduce the method of estimating $\eta_2$.
Let $\phi_i^{(3)}$, $1\le i\le 21$, be the number of subgraphs in motif $M_i^{(5)}$ that are isomorphic to motif $M_2^{(5)}$.
The value of $\phi_i^{(3)}$ is given in Table~\ref{tab:5nodephi}.
Let $\Lambda_4 =  \sum_{v\in V} \binom{d_v}{4}$.
Then, the number of all 5-node subgraphs (not necessarily induced) in $G$ isomorphic to motif $M_2^{(5)}$ is $\Lambda_4$.
Let $\Omega_3 = \{j: \phi_j^{(3)}>0\}$.
We observe that
\[
\sum_{i\in \Omega_3} \phi_{i}^{(3)} \eta_i = \Lambda_4.
\]
Since $\phi_{2}^{(3)}=1$, we estimate $\eta_2$ as
\[
\hat \eta_2 = \Lambda_4 - \sum_{i\in \Omega_3^*} \phi_{i}^{(3)} \hat\eta_i.
\]

\begin{theorem}\label{theorem: variance5nodemotifmix}
$\hat \eta_i$ is an unbiased estimator of $\eta_i$, $1\le i\le 21$.
For $i\in \Omega_1 \cup \Omega_2 = \{1,2,\ldots,21\} - \{2\}$, the variance of $\hat \eta_i$ is
\begin{equation}\label{eq:varianceMix}
\text{Var}(\hat \eta_i) = \begin{dcases}
\frac{\text{Var}(\hat \eta_i^{(1)}) \text{Var}(\hat \eta_i^{(2)})}{\text{Var}(\hat \eta_i^{(1)}) + \text{Var}(\hat \eta_i^{(2)})}, & i\in \Omega_1 \cap \Omega_2,\\
\text{Var}(\hat \eta_i^{(1)}), & i\in \Omega_1 - \Omega_2,\\
\text{Var}(\hat \eta_i^{(2)}), & i\in \Omega_2 - \Omega_1.\\
\end{dcases}
\end{equation}
For $i,j\in \Omega_1 \cup \Omega_2$ and $i\ne j$,
we compute $\text{Cov}(\hat\eta_i, \hat\eta_j)=$

\begin{equation*}
\begin{dcases}
-\sum_{l=1,2}\frac{\lambda_i^{(l)} \lambda_j^{(l)} \eta_i \eta_j}{K_l},  & i, j\in \Omega_1 \cap \Omega_2,\\
-\frac{\lambda_j^{(1)} \eta_i \eta_j}{K_1}, & i\in \Omega_1 - \Omega_2, j\in \Omega_1 \cap \Omega_2,\\
-\frac{\lambda_i^{(2)} \eta_i \eta_j}{K_2}, & i\in \Omega_1 \cap \Omega_2, j\in \Omega_2 - \Omega_1,\\
0,&i\in \Omega_1 - \Omega_2, j\in \Omega_2 - \Omega_1.
\end{dcases}
\end{equation*}
The variance of $\hat n_2$ is
\[
\text{Var} (\hat \eta_2) =\sum_{i\in \Omega_3^*} (\phi_{i}^{(3)})^2 \text{Var} (\hat\eta_i) + \sum_{i, j\in \Omega_3^*, i\ne j} \phi_{i}^{(3)} \phi_{j}^{(3)} \text{Cov}(\hat\eta_i,\hat\eta_j),
\]
where $\Omega_3^*=\Omega_3-\{2\}$.
\end{theorem}

\subsubsection{Parameter Setting}
From Theorem~\ref{theorem: variance5nodemotifmix}, we can see that the error of $\hat \eta_i$ greatly depends on the sampling budget $K_1$ for $i\in \Omega_1 - \Omega_2$.
In contrast, $K_2$ is used to guarantee the accuracy of $\hat \eta_i$, $i\in \Omega_2 - \Omega_1$.
Thus, $K_1$ and $K_2$ can be set according to the above observations.
In our experiments, we find that $p_i^{(1)}$ and $p_i^{(2)}$ have similar values.
Therefore, we set $K_1=K_2$ in this paper for simplicity.

\subsubsection{Computational Complexity}
For the T-5 sampling method, we easily extend the methods in Section~\ref{sec:samplingfunctions} to design its functions  $\text{WeightRandomVertex}(V, \rho^{(1)})$ and $\text{WeightRandomVertex}(N_v, \sigma^{(v)})$ in Algorithm~\ref{alg:T-5}. Thus, the computational complexity of T-5 sampling $K_1$ CISes is $O(|E|+K_1\log|V|)$.

For the Path-5 sampling method, we easily extend the methods in Section~\ref{sec:samplingfunctions} to design its functions  $\text{WeightRandomVertex}(V, \rho^{(2)})$ and $\text{WeightRandomVertex}(N_v, \tau^{(v)})$ in Algorithm~\ref{alg:Path-5}.
Next, we present our method of implementing $\text{WeightRandomVertex}(N_v- \{u\}, \mu^{(v, u)})$ in Algorithm~\ref{alg:Path-5}.
As alluded, we use a list $N_v[1, \ldots, d_v]$ to store the neighbors of $v$.
We store an array $ACC\_\mu^{(v)}[1, \ldots, d_v]$ in memory,
where $ACC\_\mu^{(v)}[i]$ is defined as $ACC\_\mu^{(v)}[i] = \sum_{j=1}^i (d_{N_v[j]} - 1)$, $1\le i\le d_v$.
Let $ACC\_\mu^{(v)}[0]=0$.
Let $POS_{v,u}$ be the index of $u$ in $N_v[1, \ldots, d_v]$,
i.e., $N_v[POS_{v,u}]=u$.
Then, function $\text{WeightRandomVertex}(N_v- \{u\}, \mu^{(v, u)})$ can be easily achieved by the following three steps:
\begin{itemize}
  \item Step 1: Select a number $rnd$ from
  $  \left\{1, \ldots, ACC\_\mu^{(v)}[d_v]\right\} - \left\{ACC\_\mu^{(v)}[POS_{v,u}-1] + 1, \ldots, ACC\_\mu^{(v)}[POS_{v,u}]\right\}$ at random;
  \item Step 2: Find $i$ satisfying
  \[
  ACC\_\mu^{(v)}[i-1] < rnd \le ACC\_\mu^{(v)}[i],
  \]
which can be solved by the binary search algorithm;
  \item Step 3: Return $N_v[i]$.
\end{itemize}
Its computational complexity is $O(\log d_v)$.
Therefore, the computational complexity of Path-5 sampling $K_2$ CISes is $O(|E|+K_2\log|V|)$.

\subsection{Vertex-Centric Programming Models}
In this subsection, we show MOSS-5 can be easily implemented in a vertex-centric programming model.
\subsubsection{Vertex-Centric Programming Model of T-5}
We sample $K_1$ nodes in $V$ according to $\rho^{(1)}$.
Let $k_v^{(1)}$ denote the number of times a node $v\in V$ sampled.
Thus, $\sum_{v\in V} k_v^{(1)} = K_1$.
For each node $v$, we set $k_v^{(1)}$ as its node value, and then repeat the set of five following operations $k_v^{(1)}$ times
\[
u\leftarrow \text{WeightRandomVertex}(N_v, \sigma^{(v)}),
\]
\[
w\leftarrow  \text{RandomVertex}(N_v - \{u\}),
\]
\[
r\leftarrow  \text{RandomVertex}(N_v - \{u, w\}),
\]
\[
\text{Update}(A) \text{ and then } \text{MSG}(v,*,w,r,*,A)\rightarrow u,
\]
where $A$ is the adjacent matrix of the CIS consisting of nodes $v$, $u$, $w$, $r$, and $t$,
which are the variables in Algorithm \ref{alg:T-5}, i.e., the nodes sampled at the 1-st, 2-nd, 3-rd, 4-th, and 5-th steps respectively.
Note that here $t$ and some entries in $A$ are unknown.

We process the messages that a node receives as follows:
\begin{itemize}
  \item When a node $u$ receives a message as $(v,*,w,r,*,A)$, do
\[
t \leftarrow \text{RandomVertex}(N_u - \{v\}),
\]
\[
\text{Update}(A) \text{ and then } \text{MSG}(v,u,w,r,*,A) \rightarrow t.
\]
  \item When a node $t$ receives a message as $(v,u,w,r,*,A)$, do
   \[
\text{Update}(A) \text{ and then } \text{MSG}(v,u,w,r,t,A) \rightarrow w.
\]
We send $\text{MSG}(v,u,w,r,t,A)$ to $w$ to determine whether there exists an edge between $w$ and $r$.

  \item When a node $w$ receives a message as $(v,u,w,r,t,A)$, we first $\text{Update}(A)$ and then set $m_i^{(1)}\leftarrow m_i^{(1)}+1$, where $i$ is the motif class of the CIS consisting of  $v$, $u$, $w$, $r$, and $t$.
\end{itemize}

\subsubsection{Vertex-Centric Programming Model of Path-5}
We sample $K_2$ nodes in $V$ according to $\rho^{(2)}$.
Let $k_v^{(2)}$ denote the number of times a node $v\in V$ sampled.
Thus, $\sum_{v\in V} k_v^{(2)} = K_2$.
For each node $v$, we set $k_v^{(2)}$ as its node value, and then repeat the set of five following operations $k_v^{(2)}$ times
\[
u\leftarrow \text{WeightRandomVertex}(N_v, \tau^{(v)}),
\]
\[
w\leftarrow  \text{RandomVertex}(N_v- \{u\}, \mu_w^{(v, u)}),
\]
\[
\text{Update}(A) \text{ and then } \text{MSG}(v,*,w,*,*,A)\rightarrow u,
\]
where $A$ is the adjacent matrix of the CIS consisting of nodes $v$, $u$, $w$, $r$, and $t$,
which are the variables in Algorithm \ref{alg:Path-5}, i.e., the nodes sampled at the 1-st, 2-nd, 3-rd, 4-th, and 5-th steps respectively.
Note that here $t$ and some entries in $A$ are unknown.

We process the messages that a node receives as follows:
\begin{itemize}
  \item When a node $u$ receives a message like $(v,*,w,*,*,A)$, do
\[
r \leftarrow \text{RandomVertex}(N_u - \{v\}),
\]
\[
\text{Update}(A) \text{ and then } \text{MSG}(v,u,*,r,*,A) \rightarrow w.
\]
  \item When a node $w$ receives a message like $(v,u,*,r,*,A)$, do
\[
t \leftarrow \text{RandomVertex}(N_w - \{v\}),
\]
\[
\text{Update}(A) \text{ and then } \text{MSG}(v,u,w,r,*,A) \rightarrow t.
\]
  \item When a node $t$ receives a message like $(v,u,w,r,*,A)$, do
\[
\text{Update}(A) \text{ and then } \text{MSG}(v,u,w,r,t,A) \rightarrow r.
\]
We send $\text{MSG}(v,u,w,r,t,A)$ to $r$ to determine whether there exists an edge between $v$ and $r$.
  \item When a node $r$ receives a message like $(v,u,w,r,t,A)$, we first $\text{Update}(A)$ and then set $m_i^{(2)}\leftarrow m_i^{(2)}+1$, where $i$ is the 5-node motif class of the CIS consisting of  $v$, $u$, $w$, $r$, and $t$.
\end{itemize}

\begin{figure*}[htb]
\subfigure[real values of $n_i$]{
\includegraphics[width=0.325\textwidth]{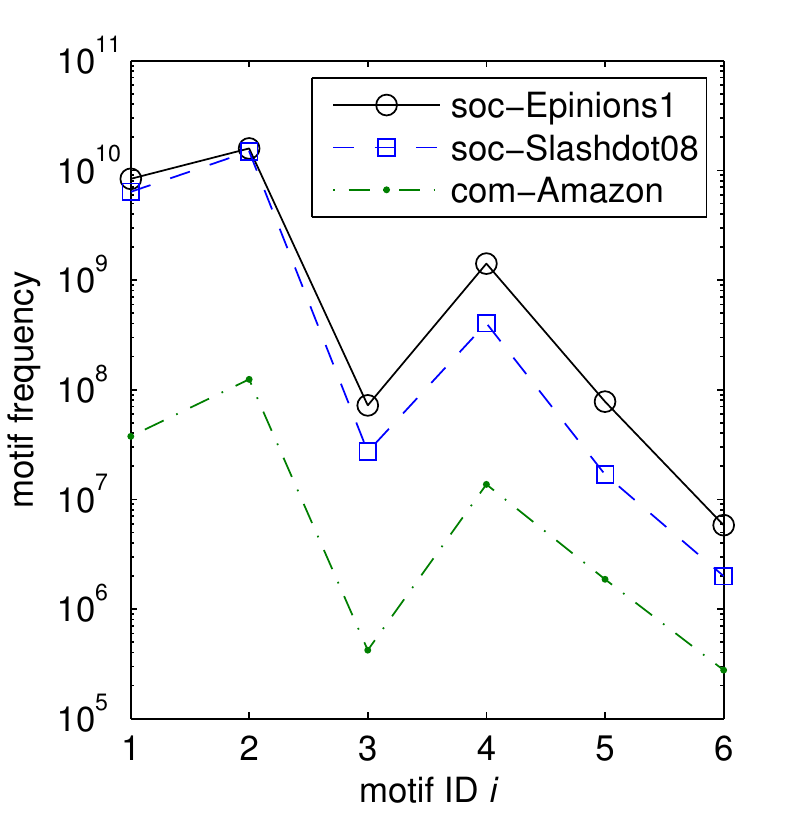}}
\subfigure[MOSS-4]{
\includegraphics[width=0.325\textwidth]{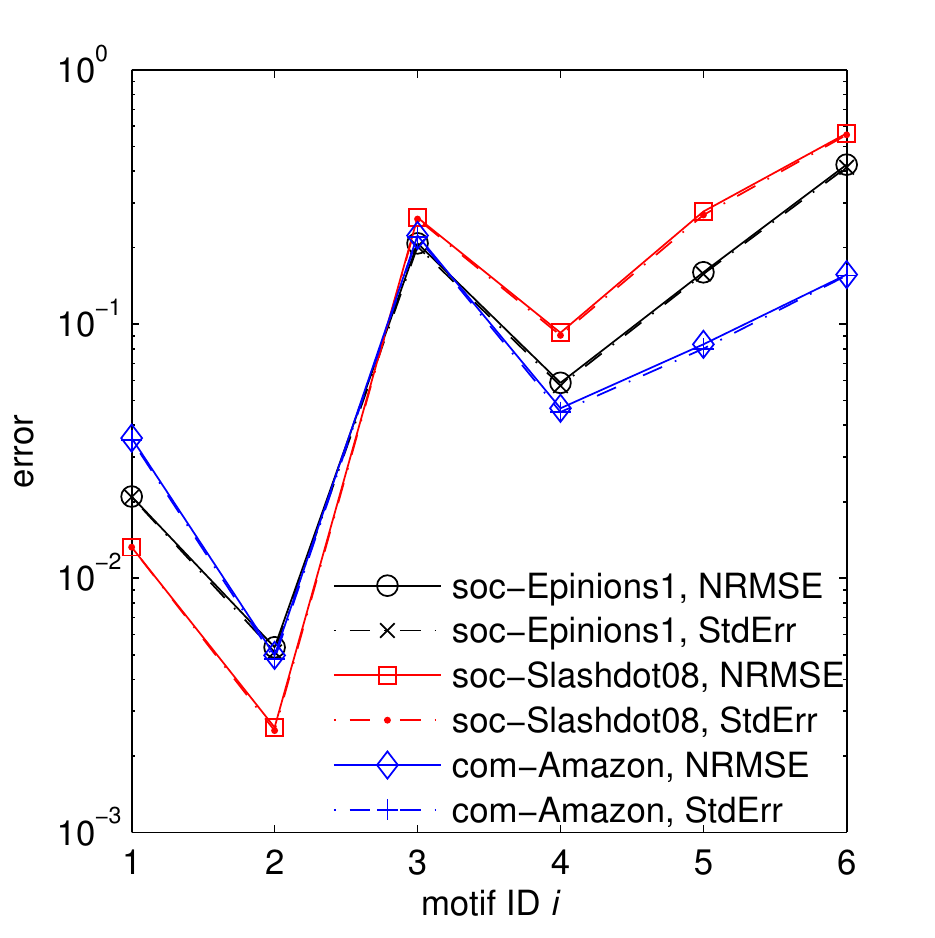}}
\subfigure[MOSS-4Min]{
\includegraphics[width=0.325\textwidth]{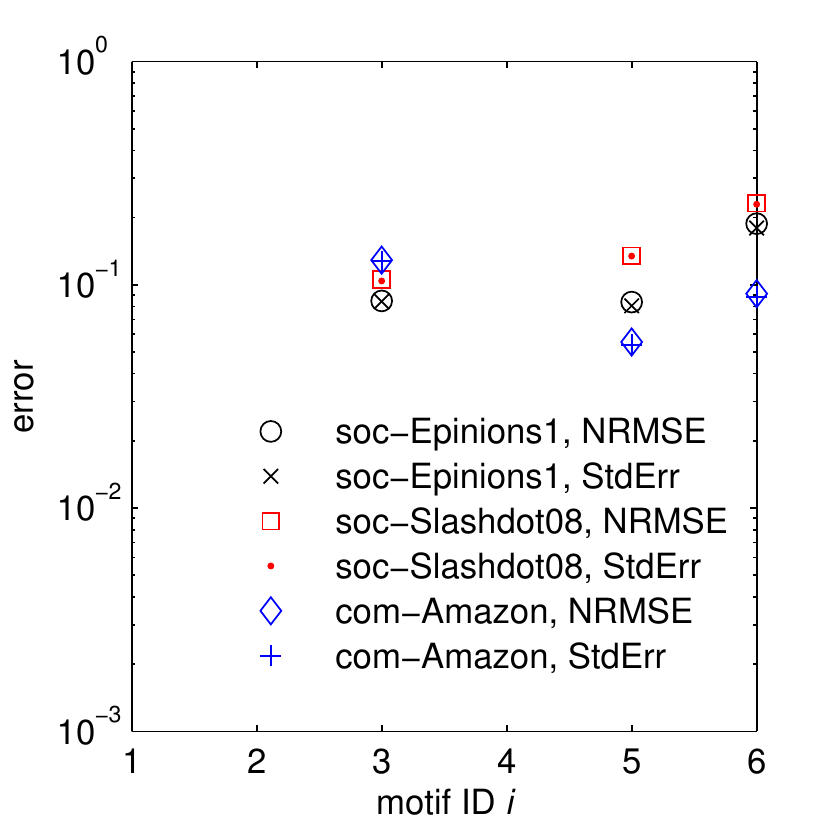}}
\caption{NRMSEs and StdErrs of the estimates of 4-node motif frequencies $n_1, \ldots, n_6$ given by MOSS-4 and MOSS-4Min, $K=1,000$ and $\check K=1,000$.}
\label{fig:MOSS4MSE}
\end{figure*}

\section{Data Evaluation} \label{sec:results}
\subsection{Datasets}
We perform our experiments on the following publicly available datasets taken from the Stanford Network Analysis Platform (SNAP)\footnote{www.snap.stanford.edu}, which are summarized in Table~\ref{tab:datasets}.
\begin{table}[htb]
\begin{center}
\caption{Graph datasets used in our experiments. ``edges" refers to the number of edges in the
undirected graph generated by discarding edge labels. ``max-degree" represents  the maximum number of edges incident to a node in the undirected graph.\label{tab:datasets}}
\begin{tabular}{|c|ccc|}
\hline
Graph&nodes&edges&max-degree\\
\hline
soc-Epinions1~\cite{Richardson2003}&75,897&405,740&3,044\\
soc-Slashdot08~\cite{LeskovecIM2009}&77,360&469,180&2,539\\
com-DBLP~\cite{YangICDM2012}&317,080&1,049,866&343\\
com-Amazon~\cite{YangICDM2012}&334,863&925,872&549\\
p2p-Gnutella08~\cite{Ripeanu2002}&6,301&20,777&97\\
ca-GrQc~\cite{LeskovecTKDD2007}&5,241&14,484&81\\
ca-CondMat~\cite{LeskovecTKDD2007}&23,133&93,439&279\\
ca-HepTh~\cite{LeskovecTKDD2007}&9,875&25,937&65\\
\hline
\end{tabular}
\end{center}
\end{table}

\subsection{Error Metric}
We study the normalized root mean square error (NRMSE) to measure the relative error of the motif frequency estimate $\hat{n}_i$ with respect to its true value $n_i$, $i=1,\dots,6$.
$\text{NRMSE}(\hat{n}_i)$ is defined as:
\[
\text{NRMSE}(\hat{n}_i)=\frac{\sqrt{\text{MSE}(\hat{n}_i)}}{n_i}, \qquad i=1,\dots,6,
\]
where $\text{MSE}(\hat{n}_i)$ is defined as
\[
\text{MSE}(\hat{n}_i)=\mathbb{E}[(\hat{n}_i-n_i)^2]=\text{Var}(\hat{n}_i)+\left(\mathbb{E}[\hat{n}_i]-n_i\right)^2.
\]
Moreover, we define a standard error (in short, StdErr) of $\hat{n}_i$ as
\[
\text{StdErr}(\hat{n}_i)=\frac{\sqrt{\text{Var}(\hat{n}_i)}}{n_i}, \qquad i=1,\dots,6.
\]
We can see that $\text{MSE}(\hat{n}_i)$ decomposes into a sum of the variance and bias of the estimator $\hat{n}_i$, both quantities are important and need to be as small as possible to achieve good estimation performance.
When $\hat{n}_i$ is an unbiased estimator of $n_i$,
then $\text{MSE}(\hat{n}_i)= \text{Var}(\hat{n}_i)$ and thus $\text{NRMSE}(\hat{n}_i)$ is equivalent to the normalized standard error of $\hat{n}_i$,
i.e., $\text{NRMSE}(\hat{n}_i)= \sqrt{\text{Var}(\hat{n}_i)}/n_i = \text{StdErr}(\hat{n}_i)$.
In our experiments, we average the estimates and calculate their NRMSEs over 1,000 runs.
Similarly, we define $\text{NRMSE}(\check{n}_i)$ and $\text{NRMSE}(\hat{\eta}_i)$ for methods MOSS-4Min and MOSS-5.
To validate the effective of our analytical error bounds,
we also compute StdErrs of MOSS-4, MOSS-4Min, and MOSS-5 based on the derived closed formula of $\text{Var}(\hat{n}_i)$, $\text{Var}(\check{n}_i)$, and $\text{Var}(\hat{\eta}_i)$.

\begin{figure}[htb]
\includegraphics[width=0.49\textwidth]{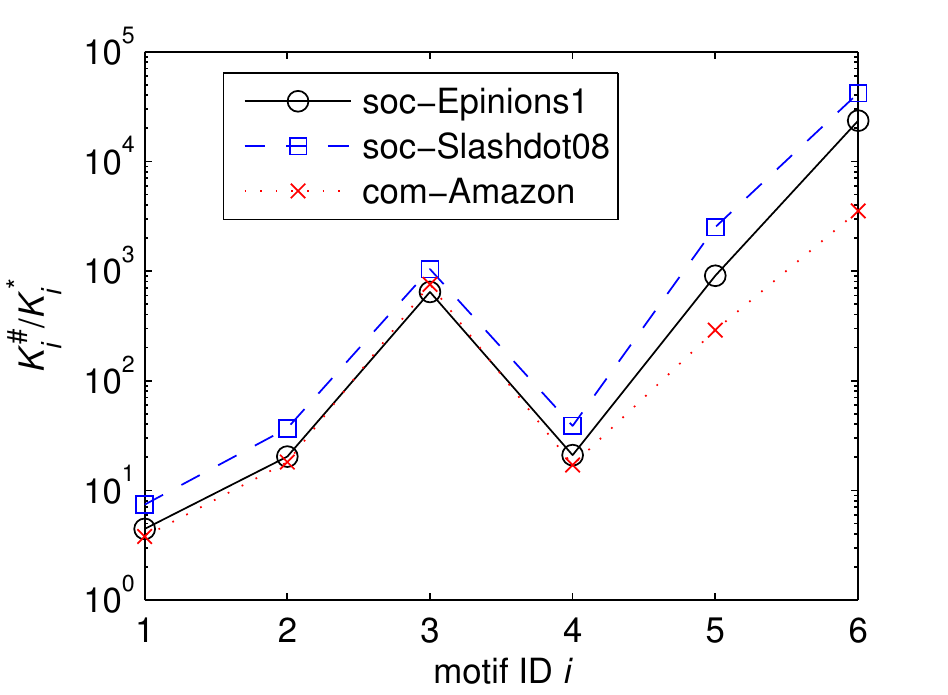}
\caption{The values of $K^\#_i/K^*_i$.}
\label{fig:cmpKratio}
\end{figure}

\begin{figure*}[htb]
\center
\includegraphics[width=0.8\textwidth]{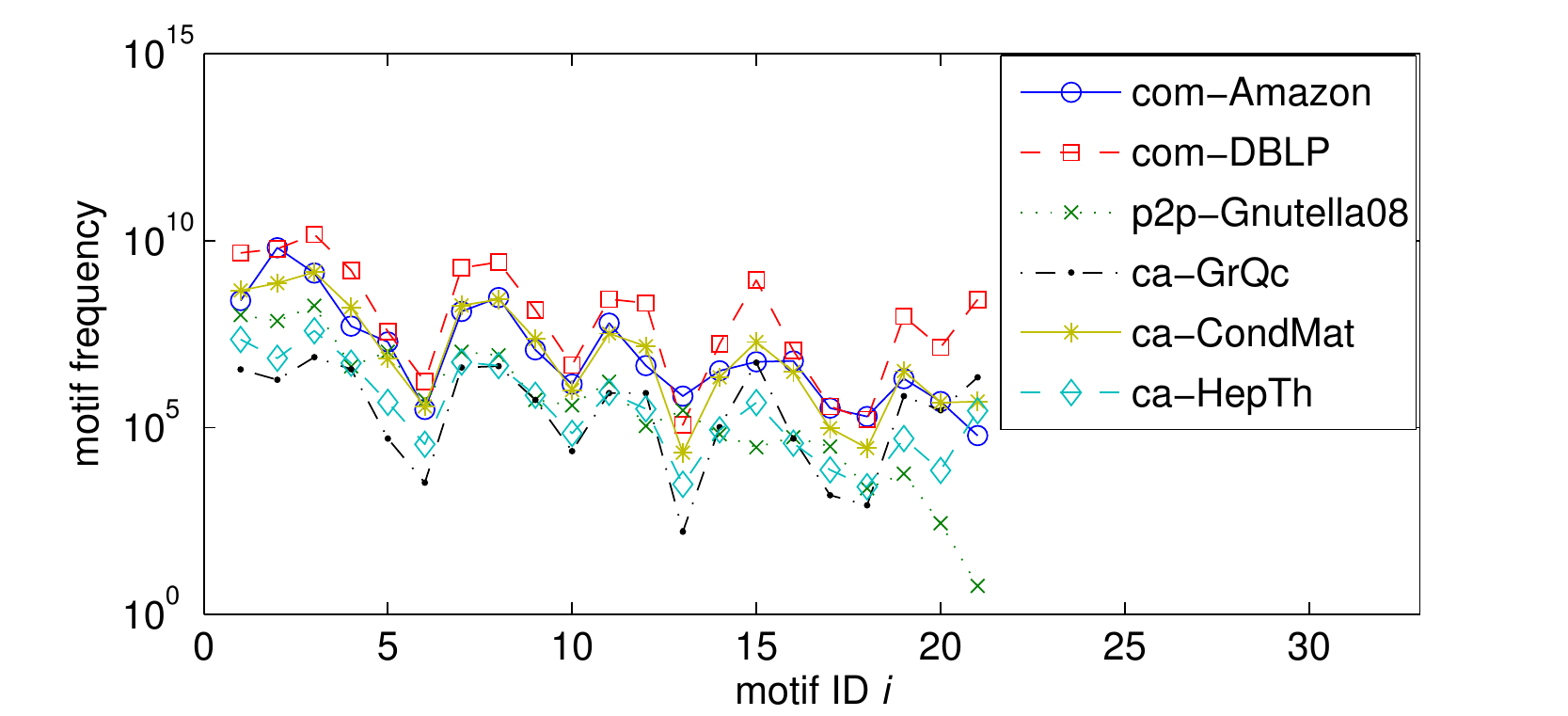}
\caption{Real values of $\eta_i$, $1\le i\le 21$.}
\label{fig:ground_stratified_5nodeCIS}
\end{figure*}

\subsection{Estimating all 4-node motifs' frequencies}
Figure~\ref{fig:MOSS4MSE}(a) shows the real values of 4-node motif frequencies $n_1^{(4)},\ldots, n_6^{(4)}$ for graphs com-Epinions1, soc-Slashdot08, and com-Amazon,
which have $2.58\times 10^{10}$, $2.17\times 10^{10}$, and $1.78\times 10^8$ 4-node CISes respectively.
We can see that the motif frequencies of $M_3^{(4)}$, $M_5^{(4)}$, and $M_6^{(4)}$ are several orders of magnitude smaller than that of the other motifs.
Fig~\ref{fig:MOSS4MSE}(b) shows the NRMSEs and StdErrs of $\hat n_1^{(4)},\ldots, \hat n_6^{(4)}$, the estimates of 4-node undirected motifs' frequencies given by MOSS-4, where we set $K=1,000$.
we can see that motifs with high frequencies exhibit larger NRMSEs and StdErrs than motifs with low frequencies.
Moreover, we observe that the StdErr of $\hat n_i^{(4)}$ almost equals to the NRMSE of $\hat n_i^{(4)}$, which is consistent to our analysis above.
Our derived error formulas indicate that the StdErr of $\hat n_i^{(4)}$ decreases linearly with the sampling budget $\sqrt K$,
which helps us to estimate the computational time required to guarantee certain accuracy for the estimate in advance.
Fig~\ref{fig:MOSS4MSE}(c) shows the NRMSEs and StdErrs of $\check n_3^{(4)}$, $\check n_5^{(4)}$, and $\check n_6^{(4)}$ given by MOSS-4MIN, where we set $\check K=1,000$.
Similarly, we see that the StdErr of $\hat n_i^{(4)}$ almost equals to the NRMSE of $\hat n_i^{(4)}$, $i=3, 5, 6$.
We compute $\frac{\text{NRMSE} (\hat n_i)} {\text{NRMSE} (\check n_i)}$ to evaluate the performance of MOSS-4Min in comparison with MOSS-4.
$\frac{\text{NRMSE} (\hat n_i)} {\text{NRMSE} (\check n_i)}$ of soc-Slashdot08 is 2.4, 1.9, and 2.3 for $i=3$, $i=5$, and $i=6$ respectively.
$\frac{\text{NRMSE} (\hat n_i)} {\text{NRMSE} (\check n_i)}$ of com-Epinions1 is 2.5, 2.0, and 2.4 for $i=3$, $i=5$, and $i=6$ respectively.
$\frac{\text{NRMSE} (\hat n_i)} {\text{NRMSE} (\check n_i)}$  of  com-Amazon is 1.7, 1.5, and 1.8 for $i=3$, $i=5$, and $i=6$ respectively.
$\frac{\text{NRMSE} (\hat n_i)} {\text{NRMSE} (\check n_i)}$  of  ca-GrQc is 1.1, 0.9, and 1.8 for $i=3$, $i=5$, and $i=6$ respectively.
We can see that MOSS-4Min exhibits a slightly improvement for ca-GrQc, so it is consistent to the analysis in Section~\ref{subsection: compareMOSS4}.
To guarantee $P(|\hat n_i -n_i|>\varepsilon n_i)<\delta$, $i=1,\ldots, 6$, we let $K_i^*$ and $K_i^\#$ denote the smallest sampling budgets that are determined by our method and the method in~\cite{JhaWWW15} respectively.
Fig.~\ref{fig:cmpKratio} shows the values of $K^\#_i/K^*_i$, where $\varepsilon=0.1$ and $\delta=0.01$.
We can see that the sampling budgets given by the method in~\cite{JhaWWW15} are several orders of magnitude larger than our method.
It indicates that the method in~\cite{JhaWWW15} does not bound the estimation error tightly and so it significantly over-estimates the sampling budget required to achieve a certain accuracy.

\begin{figure*}[htb]
\center
\subfigure[com-Amazon]{
\includegraphics[width=0.325\textwidth]{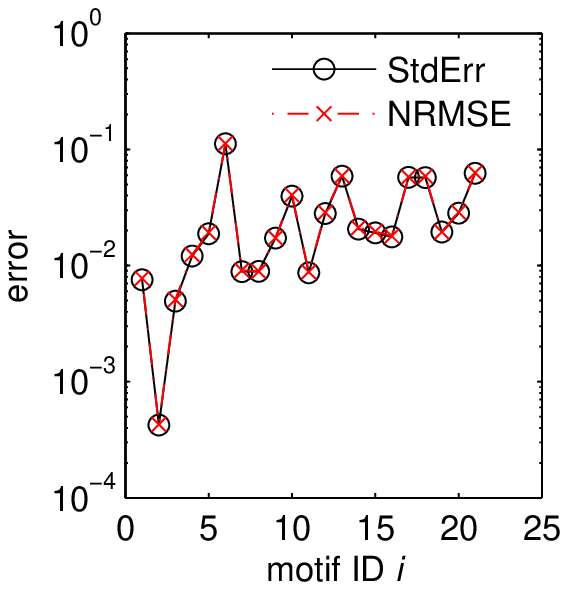}}
\subfigure[soc-DBLP]{
\includegraphics[width=0.325\textwidth]{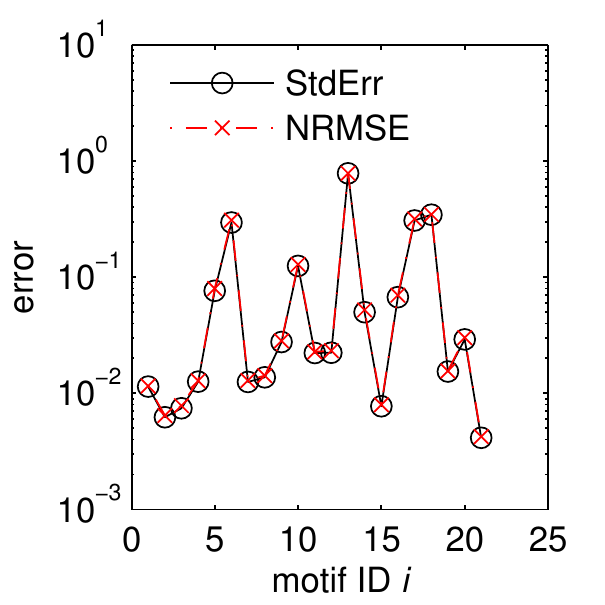}}
\subfigure[p2p-Gnutella08]{
\includegraphics[width=0.325\textwidth]{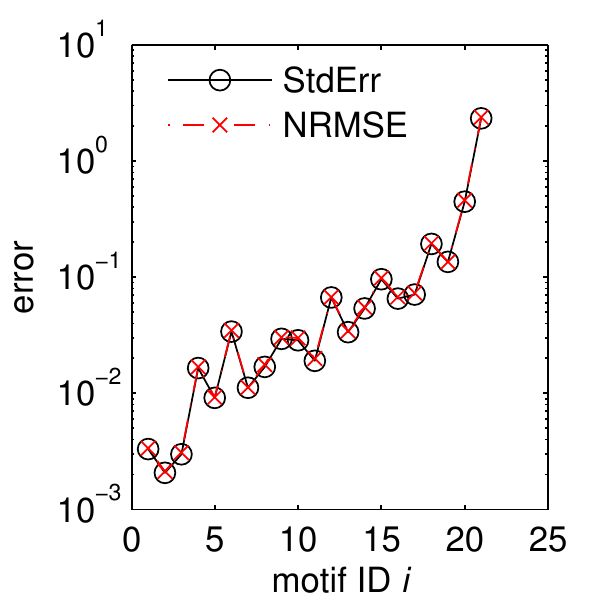}}
\subfigure[ ca-GrQc]{
\includegraphics[width=0.325\textwidth]{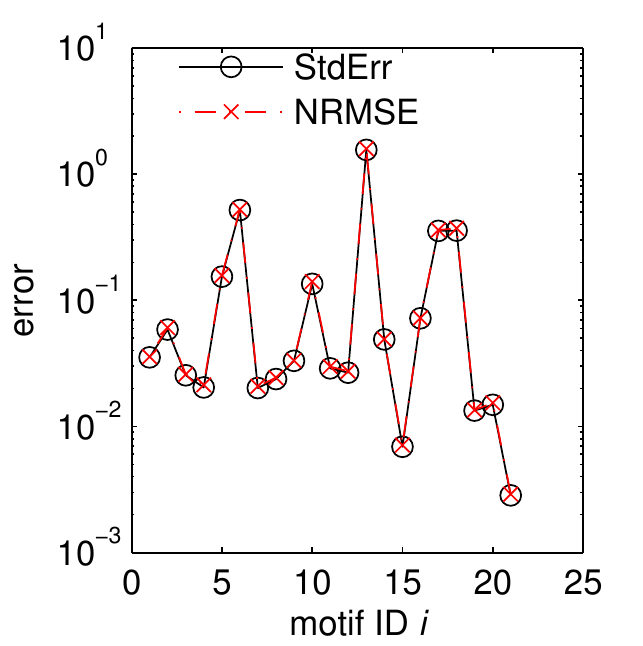}}
\subfigure[ca-CondMat]{
\includegraphics[width=0.325\textwidth]{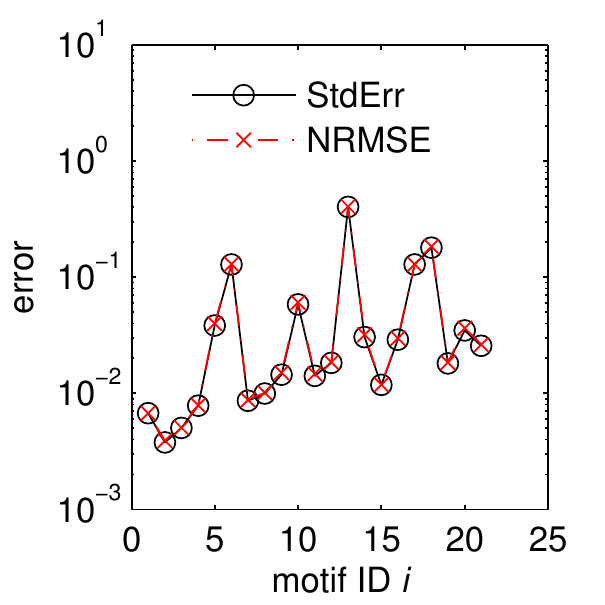}}
\subfigure[ca-HepTh]{
\includegraphics[width=0.325\textwidth]{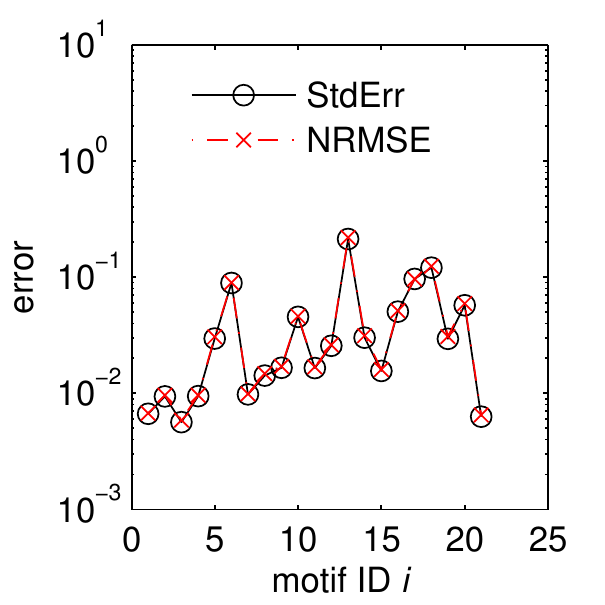}}
\caption{Real values, StdErrs, and NRMSEs of $\hat \eta_i^{(5)}$, i.e., the motif frequency estimates of $M_i^{(5)}$, $1\le i\le 21$, where $K_1 = 50,000$ and $K_2 = 50,000$.}
\label{fig:Results5nodeCIS}
\end{figure*}

\subsection{Estimating all 5-node motifs' frequencies}
\begin{figure}[htb]
\includegraphics[width=0.5\textwidth]{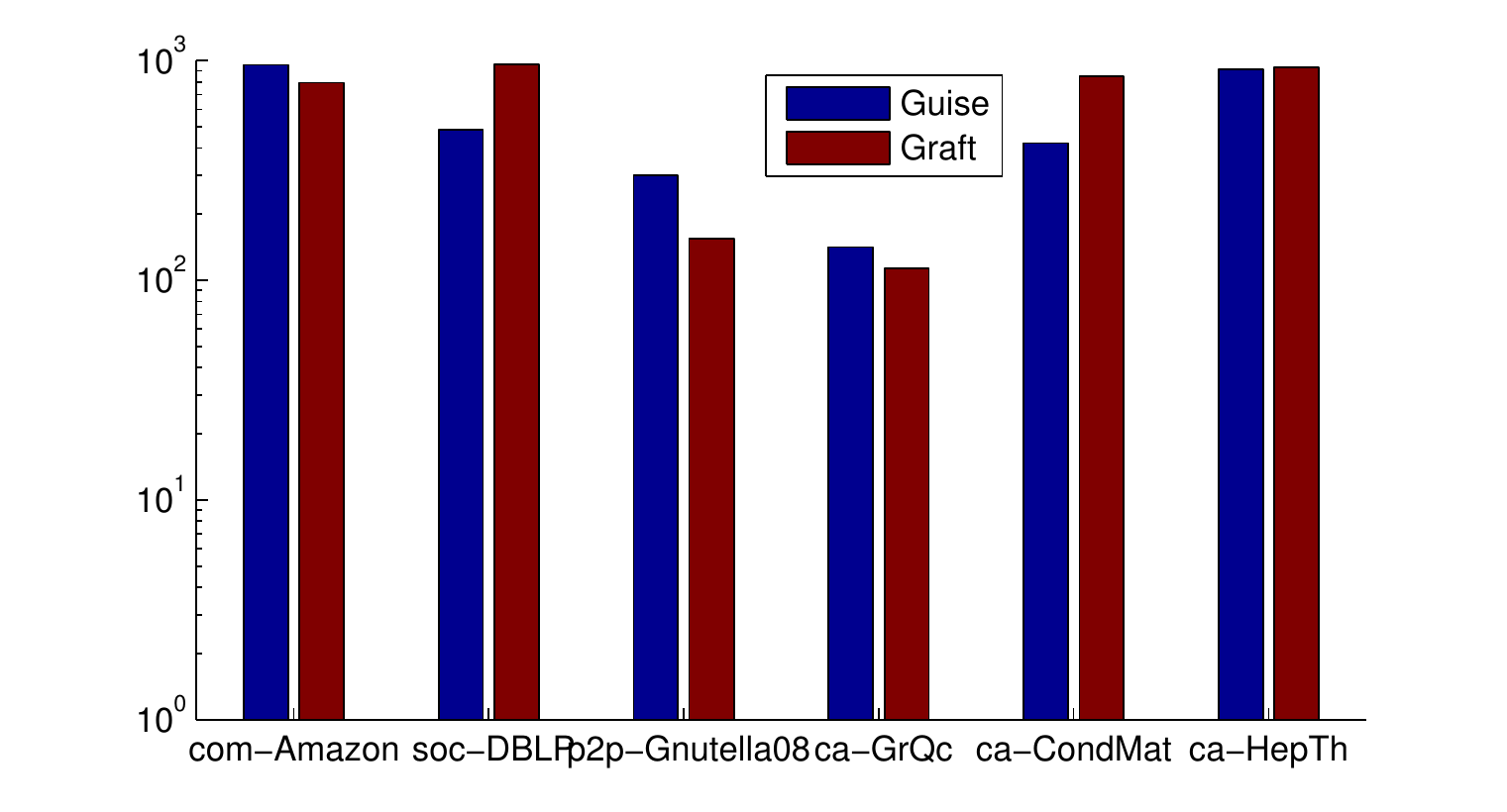}
\caption{Runtimes of the-state-of-art methods normalized with respect to runtimes of MOSS-5 for estimating 5-node motif concentrations.}
\label{fig:relativeruntime}
\end{figure}

Figure~\ref{fig:ground_stratified_5nodeCIS} shows the real values of $\eta_1,\ldots, \eta_{21}$ for graphs com-Amazon, com-DBLP, p2p-Gnutella08, ca-GrQc, ca-CondMat, and ca-HepTh,
which have $8.50\times 10^9$, $3.34\times 10^{10}$, $3.92\times 10^8$, $3.64\times 10^7$, $3.32\times 10^9$, and $8.73\times 10^7$ 5-node CISes respectively.
Fig~\ref{fig:Results5nodeCIS} shows the NRMSEs and StdErrs of $\hat \eta_1^{(5)},\ldots, \hat \eta_{21}^{(5)}$, where we set $K_1 = 50,000$ and $K_2 = 50,000$.
We can see that the StdErrs are very close to the NRMSEs.
It indicates that the derived StdErrs can be accurately used to evaluate the error of our estimates given by MOSS-5.
To the best of our knowledge, MOSS-5 is the first to provide a simple and accurate formula for analyzing estimation errors of 5-node motif frequencies.
The results show that the NRMSEs of all 5-node motifs are smaller than 0.1 for com-Amazon, which is larger than the other graphs studied in this paper.
For the other graphs, most 5-node motifs' NRMSEs are smaller than 0.1.
The NRMSE of $\hat \eta_{21}^{(5)}$ is larger than 1 for p2p-Gnutella08,
and the NRMSE of $\hat \eta_{21}^{(5)}$ is larger than 1 for ca-GrQc.
We observe that p2p-Gnutella08 has only several CISes isomorphic to $M_{21}^{(5)}$,
and p2p-Gnutella08 has no more than 200 CISes isomorphic to $M_{13}^{(5)}$.
It is very challenging to observe and count these rare motifs for sampling based methods.
Most previous work focuses on estimating 5-node motif concentrations, which is defines as $\omega_i = \frac{\eta_i}{\sum_{j=1}^{21} \eta_j}$, $i=1,\ldots,21$.
We run MOSS-5, state-of-the-arts methods Guise~\cite{Bhuiyan2012} and Graft~\cite{GRAFT2012} over all above graphs and increase their sampling budgets until the estimation errors of motif concentrations are within 10\%.
Fig.~\ref{fig:relativeruntime} shows the runtimes of Graft and Guise normalized with respect to the runtimes of MOSS-5.
We can see that our method MOSS-5 is 2 to 3 orders of magnitude faster than Graft and Guise.

\section{Related Work} \label{sec:related}
In this paper, we study the problem of computing 4- and 5-node motifs' frequencies for \emph{a single large graph},
which is much different from the problem of computing the number of subgraph patterns appearing in \emph{a large set of graphs} studied in~\cite{Hasan2009}.
Recently, a lot of efforts has been devoted to design sampling methods for computing a large graph's motif concentrations~\cite{Kashtan2004,Wernicke2006,OmidiGenes2009,Bhuiyan2012,GRAFT2012,TKDDWang2014}.
However, these methods fail to compute motif frequencies, which is more fundamental than motif concentrations.
Alon et al.~\cite{Alon1995} propose the color-coding method to reduce the computational cost of counting subgraphs.
Color coding reduces the computations by coloring nodes randomly and enumerating only colorful CISes (i.e., CISes that are consisted of nodes with distinct colors),
but~\cite{JhaKDD2015} reaveals that the color-coding method is not scalable and is hindered by the sheer number of colorful CISes.
\cite{TsourakakisKDD2009,PavanyVLDB2013,JhaKDD2013,AhmedKDD2014} develop sampling methods to estimate the number of triangles of static and dynamic graphs.
Jha et al.~\cite{JhaWWW15} develop sampling methods to estimate 4-node undirected motifs' frequencies.
However their methods are edge centric methods, which cannot be easily applied to current vertex centric graph computing systems such as GraphLab~\cite{LowPVLDB2012} and GraphChi~\cite{KyrolaOSDI12}.
Moreover, their methods fail to sample and count 5-node motifs.

\section{Conclusions} \label{sec:conclusions}
We develop computationally efficient sampling methods MOSS-4 and MOSS-5 to estimate the frequencies of all 4- and 5-node motifs.
Compared MOSS-4, MOSS-4Min is better to characterize rare motifs.
All these methods provide unbiased estimators of motif frequencies, and we derive simple and exact formulas for the variances of the estimators.
Meanwhile, we conduct experiments on a variety of publicly available datasets,
and experimental results show that our methods significantly outperform state-of-the-art methods.

\section*{Acknowledgment}
This work was supported in part by the National Natural Science Foundation of China (61103240, 61103241, 61221063, 61221063, 91118005, U1301254),
the 111 International Collaboration Program of China, 863 High Tech Development Plan (2012AA011003), the Prospective Research Project
on Future Networks of Jiangsu Future Networks Innovation Institute, and
the Application Foundation Research Program of SuZhou (SYG201311).

\section*{Appendix}

\subsection*{Proof of Theorem~\ref{theorem: MOSS4samplingweight}}
\begin{figure}[htb]
\center
\includegraphics[width=0.49\textwidth]{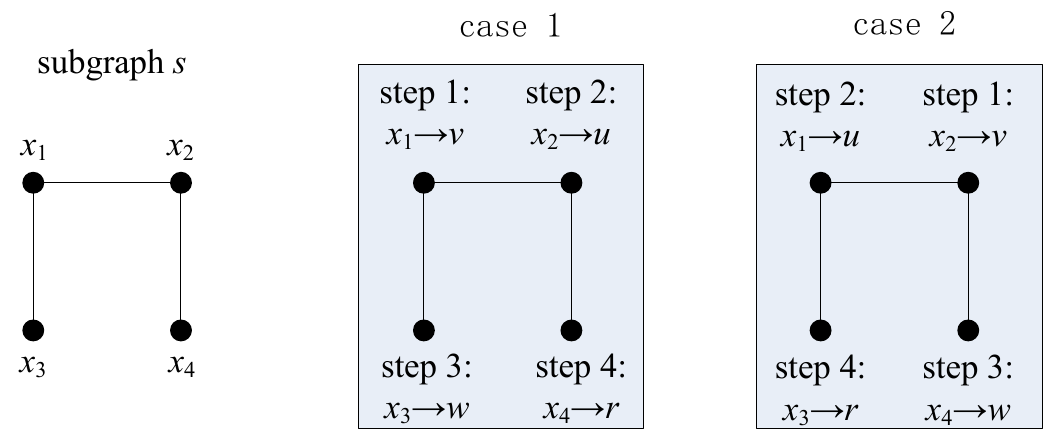}
\caption{The ways of MOSS-4 sampling a subgraph $s$ isomorphic to motif $M_1^{(4)}$, where $v$, $u$, $w$, and $r$ are the variables in Algorithm \ref{alg:MOSS4}, i.e., the nodes sampled at the 1-st, 2-nd, 3-rd, and 4-th steps respectively.}
\label{fig:example4nodes}
\end{figure}

As shown in Fig.~\ref{fig:example4nodes},
we find that there exist two ways to sample a subgraph isomorphic to motif $M_1^{(4)}$ by MOSS-4.
Each one happens with probability $\pi_v\times \sigma_u^{(v)} \times \frac{1}{d_v-1} \times \frac{1}{d_u-1}=\frac{1}{\Gamma}$.
For a 4-node CIS $s$ isomorphic to motif $M_i^{(4)}$,
$s$ has $\varphi_i^{(1)}$ subgraphs isomorphic to motif $M_1^{(4)}$, $1\le i\le 6$.
Thus, there exist $2\varphi_i^{(1)}$ ways to sample $s$ by MOSS-4,
and the probability of MOSS-4 sampling $s$ is $\frac{2\varphi_i^{(1)}}{\Gamma}$.

\subsection*{Proof of Theorem~\ref{theorem: varianceMOSS4}}
For $i\in \{1,3,4,5,6\}$ and $1\le k\le K$,
we have
\[
P(M^{(4)}(s_k)=i) = \sum_{s\in C^{(4)}} P(s_k = s) \mathbf{1}(M^{(4)}(s)=i) = p_i n_i.
\]
$s_1, \ldots, s_K$ are sampled independently,
so the random variable $m_i$ follows the binomial distribution with parameters $K$ and $p_i n_i$.
Formally, we have
\[
P(m_i = x)= \binom{K}{x} (p_i n_i)^x (1-p_i n_i)^{K-x}, \quad x=0, 1, \ldots, K.
\]
Then, the expectation and variance of $m_i$ are
\[
\mathbb{E} (m_i) = K p_i n_i,
\]
and
\[
\text{Var} (m_i) = K p_i n_i(1 - p_i n_i).
\]
Therefore, the expectation and variance of $\hat n_i$ are computed as
\begin{equation}\label{eq:MOSS4unbiased_ni}
\mathbb{E} (\hat n_i) = \mathbb{E} \left(\frac{m_i}{K p_i}\right) = \frac{\mathbb{E}(m_i)}{K p_i} = n_i,
\end{equation}
and
\begin{equation}\label{eq:MOSS4variance_ni}
\text{Var} (\hat n_i) = \text{Var}\left(\frac{m_i}{K p_i}\right) = \frac{\text{Var}(m_i)}{K^2 p_i^2} = \frac{n_i}{K} \left(\frac{1}{p_i} - n_i\right).
\end{equation}

From~(\ref{eq:MOSS4unbiased_ni}), we compute the expectation of $\hat n_2$ as
\begin{equation*}
\begin{split}
\mathbb{E} (\hat n_2) &= \mathbb{E} (\Lambda_3 - \hat n_4 - 2 \hat n_5 - 4 \hat n_6)\\
&=\Lambda_3 - \mathbb{E} (\hat n_4) - 2 \mathbb{E} (\hat n_5) - 4 \mathbb{E} (\hat n_6)\\
&=\Lambda_3 - n_4 - 2 n_5 - 4 n_6\\
&=n_2.
\end{split}
\end{equation*}
The last equation holds because of~(\ref{eq:n2}).
To derive the variance of $\hat n_2$, we first compute the covariance of $\hat n_i$ and $\hat n_j$, where $i\ne j$ and $i, j\in \{1,3,4,5,6\}$.
That is,
\begin{eqnarray}
&&\text{Cov}(\hat n_i, \hat n_j)\nonumber\\
&=& \text{Cov} \left(\frac{m_i}{K p_i}, \frac{m_j}{K p_j}\right)\nonumber \\
&=& \frac{\text{Cov} (\sum_{k=1}^{K} \mathbf{1}(M^{(4)}(s_k)=i), \sum_{l=1}^{K} \mathbf{1}(M^{(4)}(s_l)=j))}{K^2 p_i p_j}\nonumber \\
&=& \frac{ \sum_{k=1}^{K} \sum_{l=1}^{K} \text{Cov}(\mathbf{1}(M^{(4)}(s_k)=i), \mathbf{1}(M^{(4)}(s_l)=j))}{K^2 p_i p_j}\nonumber \\
&=& \frac{ \sum_{k=1}^{K} \text{Cov}(\mathbf{1}(M^{(4)}(s_k)=i), \mathbf{1}(M^{(4)}(s_k)=j))}{K^2 p_i p_j}\nonumber \\
&=& -\frac{n_i n_j}{K}. \label{eq:MOSS4covariance}
\end{eqnarray}
In the derivation above, we use
\begin{equation*}
\begin{split}
&\text{Cov}(\mathbf{1}(M^{(4)}(s_k)=i), \mathbf{1}(M^{(4)}(s_k)=j))\\
&= \text{E}(\mathbf{1}(M^{(4)}(s_k)=i)\mathbf{1}(M^{(4)}(s_k)=j))\\
&\quad - \text{E}(\mathbf{1}(M^{(4)}(s_k)=i)) \text{E}(\mathbf{1}(M^{(4)}(s_k)=j))\\
&=0-p_i n_i p_j n_j\\
&= -p_i  p_j  n_i n_j,
\end{split}
\end{equation*}
and $\text{Cov}(\mathbf{1}(M^{(4)}(s_k)=i), \mathbf{1}(M^{(4)}(s_l)=j)) = 0$ when $k\ne l$.
Finally, we compute the variance of $\hat n_2$ as
\begin{equation*}
\begin{split}
\text{Var} (\hat n_2) &= \text{Var} (\Lambda_3 - \hat n_4 - 2 \hat n_5 - 4 \hat n_6)\\
&= \text{Var} (\hat n_4 + 2 \hat n_5 + 4 \hat n_6)\\
&= \text{Var} (\hat n_4) + 4\text{Var} (\hat n_5) + 16\text{Var} (\hat n_6)\\
&\quad + 4\text{Cov} (\hat n_4 \hat n_5) + 8 \text{Cov} (\hat n_4 \hat n_6) + 16 \text{Cov} (\hat n_5 \hat n_6).
\end{split}
\end{equation*}
Using~(\ref{eq:MOSS4variance_ni}) and~(\ref{eq:MOSS4covariance}), then we have
\[
\text{Var} (\hat n_2) = \frac{1}{K} \left(\frac{n_4}{p_4} + \frac{4 n_5}{p_5} + \frac{16 n_6}{p_6} -(n_4 + 2 n_5 + 4 n_6)^2\right).
\]

\subsection*{Proof of Theorem~\ref{theorem: MOSS4Minsamplingweight}}
Let $\check \varphi^{(1)} (s)$ denote the number of ways to sample a 4-node CIS $s$ by MOSS-4Min.
Then, we have $\check p(s) =\check \varphi^{(1)} (s) \times \check \pi_v \times \check \sigma_u^{(v)} \times \frac{1}{d_{v,u}} \times \frac{1}{d_{u,v}}=\check \varphi^{(1)} (s) \check \Gamma^{-1}$.
We compute $\check \varphi^{(1)} (s_1) = 2$, $\check \varphi^{(1)} (s_2) = 2$, and $\check \varphi^{(1)} (s_3) = 6$ for cases $s_1\in C_3^{(4)}$,
$s_2\in C_5^{(4)}$, and $s_3\in C_6^{(4)}$ respectively.

\subsection*{Proof of Theorem~\ref{theorem: Tsamplersamplingweight}}
\begin{figure}[htb]
\center
\includegraphics[width=0.49\textwidth]{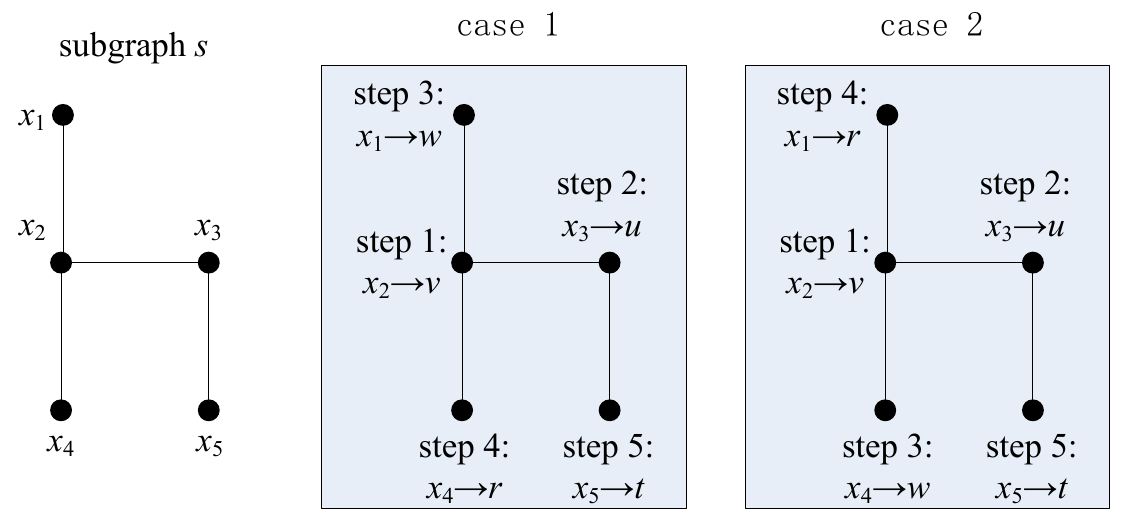}
\caption{The ways of T-5 sampling a subgraph $s$ isomorphic to motif $M_3^{(5)}$, where $v$, $u$, $w$, $r$, and $t$ are the variables in Algorithm \ref{alg:T-5}, i.e., the nodes sampled at the 1-st, 2-nd, 3-rd, 4-th, and 5-th steps respectively.}
\label{fig:example5Tsampling}
\end{figure}

As shown in Fig.~\ref{fig:example5Tsampling},
we find that there exist two ways to sample a subgraph isomorphic to motif $M_3^{(5)}$ by T-5.
Each one happens with probability $\rho_v^{(1)}\times \sigma_u^{(v)} \times \frac{1}{d_v-1} \times \frac{1}{d_v-2} \times \frac{1}{d_u-1} =\frac{1}{\Gamma^{(1)}}$.
For a 5-node CIS $s$ isomorphic to motif $M_i^{(5)}$,
$s$ has $\phi_i^{(1)}$ subgraphs isomorphic to motif $M_3^{(5)}$, $1\le i\le 21$.
Therefore, the probability of sampling $s$ is $\frac{2\phi_i^{(1)}}{\Gamma^{(1)}}$.

\subsection*{Proof of Theorem~\ref{theorem: varianceT-5}}
For $i\in \Omega_1$ and $1\le k\le K_1$, we have
\begin{equation*}
\begin{split}
P(M^{(5)}(s_k^{(1)})=i) &= \sum_{s\in C^{(5)}} P(s_k^{(1)} = s) \mathbf{1}(M^{(5)}(s_k^{(1)})=i)\\
&= p_i^{(1)} \eta_i.
\end{split}
\end{equation*}
Since $s_1^{(1)}, \ldots, s_{K_1}^{(1)}$ are sampled independently,
the random variable $m_i^{(1)}$ follows the binomial distribution with parameters $K_1$ and $p_i^{(1)} \eta_i$.
Then, the expectation and variance of $m_i^{(1)}$ are
\[
\mathbb{E} (m_i^{(1)}) = K_1 p_i^{(1)} \eta_i,
\]
and
\begin{equation}\label{eq:variance_T-5}
\text{Var} (m_i^{(1)}) = K_1 p_i^{(1)} \eta_i(1 - p_i^{(1)} \eta_i).
\end{equation}
Therefore, the expectation and variance of $\hat \eta_i^{(1)}$ are computed as
\begin{equation*}
\mathbb{E} (\hat \eta_i^{(1)}) = \mathbb{E} \left(\frac{m_i^{(1)}}{K_1 p_i^{(1)}}\right) = \frac{\mathbb{E}(m_i^{(1)})}{K_1 p_i^{(1)}} = \eta_i,
\end{equation*}
and
\begin{equation*}
\text{Var} (\hat \eta_i) = \text{Var}\left(\frac{m_i^{(1)}}{K_1 p_i^{(1)}}\right) = \frac{\eta_i}{K_1} \left(\frac{1}{p_i^{(1)}} - \eta_i\right).
\end{equation*}

For $i\ne j$ and $i, j\in \Omega_1$,
the covariance of $\hat \eta_i^{(1)}$ and $\hat \eta_j^{(1)}$ is
\begin{eqnarray}
&&\text{Cov}(\hat \eta_i^{(1)}, \hat \eta_j^{(1)})\nonumber\\
&=& \text{Cov} \left(\frac{m_i^{(1)}}{K_1 p_i^{(1)}}, \frac{m_j^{(1)}}{K_1 p_j^{(1)}}\right)\nonumber \\
&=& \frac{\text{Cov} (\sum_{k=1}^{K_1} \mathbf{1}(M^{(5)}(s_k^{(1)})=i), \sum_{l=1}^{K_1} \mathbf{1}(M^{(5)}(s_l^{(1)})=j))}{K_1^2 p_i^{(1)} p_j^{(1)}}\nonumber \\
&=& \frac{ \sum_{k=1}^{K_1} \sum_{l=1}^{K_1} \text{Cov}(\mathbf{1}(M^{(5)}(s_k^{(1)})=i), \mathbf{1}(M^{(5)}(s_l^{(1)})=j))}{K_1^2 p_i^{(1)} p_j^{(1)}}\nonumber \\
&=& \frac{ \sum_{k=1}^{K_1} \text{Cov}(\mathbf{1}(M^{(5)}(s_k^{(1)})=i), \mathbf{1}(M^{(5)}(s_k^{(1)})=j))}{K_1^2 p_i^{(1)} p_j^{(1)}}\nonumber \\
&=& -\frac{\eta_i \eta_j}{K_1}.\nonumber
\end{eqnarray}
In the derivation above, we use
\[
\text{Cov}(\mathbf{1}(M^{(5)}(s_k^{(1)})=i), \mathbf{1}(M^{(5)}(s_l^{(1)})=j)) = 0, \quad k\ne l,
\]
and
\begin{equation*}
\begin{split}
&\text{Cov}(\mathbf{1}(M^{(5)}(s_k^{(1)})=i), \mathbf{1}(M^{(5)}(s_k^{(1)})=j))\\
&= \text{E}(\mathbf{1}(M^{(5)}(s_k^{(1)})=i)\mathbf{1}(M^{(5)}(s_k^{(1)})=j))\\
&\quad - \text{E}(\mathbf{1}(M^{(5)}(s_k^{(1)})=i)) \text{E}(\mathbf{1}(M^{(5)}(s_k^{(1)})=j))\\
&=0-p_i^{(1)} \eta_i p^{(1)}_j \eta_j\\
&= -p_i^{(1)}  p_j^{(1)}  \eta_i \eta_j.
\end{split}
\end{equation*}

\subsection*{Proof of Theorem~\ref{theorem: Path-5samplingweight}}
\begin{figure}[htb]
\center
\includegraphics[width=0.4\textwidth]{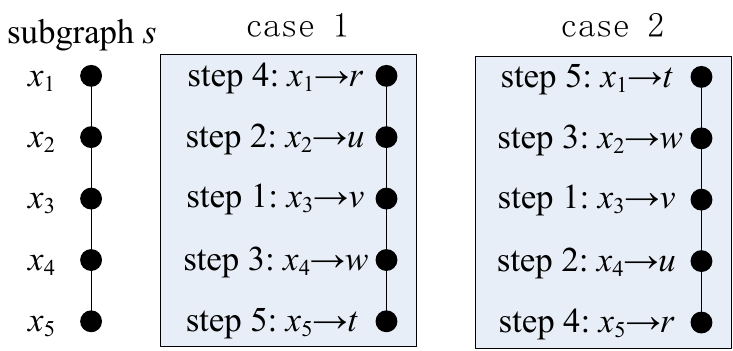}
\caption{The ways of Path-5 sampling a subgraph $s$ isomorphic to motif $M_1^{(5)}$, where $v$, $u$, $w$, $r$, and $t$ are the variables in Algorithm \ref{alg:Path-5}, i.e.,  the nodes sampled at the 1-st, 2-nd, 3-rd, 4-th, and 5-th steps respectively.}
\label{fig:example5pathsampling}
\end{figure}

As shown in Fig.~\ref{fig:example5pathsampling},
we can see that there exist two ways to sample a subgraph isomorphic to motif $M_1^{(5)}$ by our Path-5 sampling method.
Each one happens with probability $\rho_v^{(2)}\times \tau_u^{(v)} \times \mu_w^{(v, u)} \times \frac{1}{d_u-1} \times \frac{1}{d_w-1}=\frac{1}{\Gamma^{(2)}}$.
For a 5-node CIS $s$ isomorphic to the $i$-th 5-node motif,
$s$ has $\phi_i^{(2)}$ subgraphs isomorphic to motif $M_1^{(5)}$, $1\le i\le 21$.
Thus, the probability of Path-5 sampling $s$ is $\frac{2\phi_i^{(2)}}{\Gamma^{(2)}}$.

\subsection*{Proof of Theorem~\ref{theorem: variancePath-5}}
For $i\in \Omega_2$ and $1\le k\le K_2$, we have
\begin{equation*}
\begin{split}
P(M^{(5)}(s_k^{(2)})=i) &= \sum_{s\in C^{(5)}} P(s_k^{(2)} = s) \mathbf{1}(M^{(5)}(s_k^{(2)})=i)\\
&= p_i^{(2)} \eta_i.
\end{split}
\end{equation*}
$s_1^{(2)}, \ldots, s_{K_2}^{(2)}$ are sampled independently,
therefore the random variable $m_i^{(2)}$ follows the binomial distribution with parameters $K_2$ and $p_i^{(2)} \eta_i$.
Then, the expectation and variance of $m_i^{(2)}$ are
\[
\mathbb{E} (m_i^{(2)}) = K_2 p_i^{(2)} \eta_i,
\]
and
\[
\text{Var} (m_i^{(2)}) = K_2 p_i^{(2)} \eta_i(1 - p_i^{(2)} \eta_i).
\]
Thus, the expectation and variance of $\hat \eta_i^{(2)}$ are computed as
\begin{equation*}
\mathbb{E} (\hat \eta_i^{(2)}) = \mathbb{E} \left(\frac{m_i^{(2)}}{K_2 p_i^{(2)}}\right) = \frac{\mathbb{E}(m_i^{(2)})}{K_2 p_i^{(2)}} = \eta_i,
\end{equation*}
and
\begin{equation*}
\text{Var} (\hat \eta_i) = \text{Var}\left(\frac{m_i^{(2)}}{K_2 p_i^{(2)}}\right) = \frac{\eta_i}{K_2} \left(\frac{1}{p_i^{(2)}} - \eta_i\right).
\end{equation*}

For $i\ne j$ and $i, j\in \Omega_2$,
the covariance of $\hat \eta_i^{(2)}$ and $\hat \eta_j^{(2)}$ is
\begin{eqnarray}
&&\text{Cov}(\hat \eta_i^{(2)}, \hat \eta_j^{(2)})\nonumber\\
&=& \text{Cov} \left(\frac{m_i^{(2)}}{K_2 p_i^{(2)}}, \frac{m_j^{(2)}}{K_2 p_j^{(2)}}\right)\nonumber \\
&=& \frac{\text{Cov} (\sum_{k=1}^{K_2} \mathbf{1}(M^{(5)}(s_k^{(2)})=i), \sum_{l=1}^{K_2} \mathbf{1}(M^{(5)}(s_l^{(2)})=j))}{K_2^2 p_i^{(2)} p_j^{(2)}}\nonumber \\
&=& \frac{ \sum_{k=1}^{K_2} \sum_{l=1}^{K_2} \text{Cov}(\mathbf{1}(M^{(5)}(s_k^{(2)})=i), \mathbf{1}(M^{(5)}(s_l^{(2)})=j))}{K_2^2 p_i^{(2)} p_j^{(2)}}\nonumber \\
&=& \frac{ \sum_{k=1}^{K_2} \text{Cov}(\mathbf{1}(M^{(5)}(s_k^{(2)})=i), \mathbf{1}(M^{(5)}(s_k^{(2)})=j))}{K_2^2 p_i^{(2)} p_j^{(2)}}\nonumber \\
&=& -\frac{\eta_i \eta_j}{K_2}.\nonumber
\end{eqnarray}
In the derivation above, we use
\begin{equation*}
\begin{split}
&\text{Cov}\left(\mathbf{1}(M^{(5)}(s_k^{(2)})=i), \mathbf{1}(M^{(5)}(s_k^{(2)})=j)\right)\\
&= \mathbb{E}(\mathbf{1}(M^{(5)}(s_k^{(2)})=i)\mathbf{1}(M^{(5)}(s_k^{(2)})=j))\\
&- \mathbb{E}(\mathbf{1}(M^{(5)}(s_k^{(2)})=i)) \mathbb{E}(\mathbf{1}(M^{(5)}(s_k^{(2)})=j))\\
&=0-p_i^{(2)} \eta_i p^{(2)}_j \eta_j\\
&= -p_i^{(2)}  p_j^{(2)}  \eta_i \eta_j,
\end{split}
\end{equation*}
and $\text{Cov}(\mathbf{1}(M^{(5)}(s_k^{(2)})=i), \mathbf{1}(M^{(5)}(s_l^{(2)})=j)) = 0$, $k\ne l$.

\subsection*{Proof of Theorem~\ref{theorem: variance5nodemotifmix}}
For $i\in \Omega_1 \cup \Omega_2$,
Theorems~\ref{theorem: varianceT-5} and~\ref{theorem: variancePath-5} tell us that
$\eta_i^{(1)}$ and $\eta_i^{(2)}$ are unbiased estimators of $\eta_i^{(1)}$, and they are independent.
Moreover, $\lambda_i^{(1)} + \lambda_i^{(2)} =1$.
Therefore, we easily find that $\hat \eta_i$ is also an unbiased estimator of $\eta_i^{(1)}$, and its variance is~(\ref{eq:varianceMix}).
Next, we study the expectation and variance of $\hat \eta_2$.
The expectation of $\hat \eta_2$ is
 \[
\mathbb{E} (\hat \eta_2) = \Lambda_4 - \sum_{i\in \Omega_3^*} \phi_{i}^{(3)} \mathbb{E} (\hat\eta_i) =  \Lambda_4 - \sum_{i\in \Omega_3^*} \phi_{i}^{(3)} \eta_i = \eta_2.
\]
Next, we compute the covariance of $\hat\eta_i$ and $\hat\eta_j$ for $i,j\in \Omega_1 \cup \Omega_2$ and $i\ne j$.
For any $i,j\in \Omega_1 \cup \Omega_2$, we have
$\text{Cov}(\hat\eta_i^{(1)}, \hat\eta_j^{(2)}) = 0$ because $\hat\eta_i^{(1)}$ and $\hat\eta_j^{(2)}$ are independent.
Thus, we have $\text{Cov}(\hat\eta_i, \hat\eta_j) = \text{Cov}(\hat\eta_i^{(1)}, \hat\eta_j^{(2)}) = 0$
when $i\in \Omega_1 - \Omega_2$ and $j\in \Omega_2 - \Omega_1$.
When $i\in \Omega_1 - \Omega_2$ and $j\in \Omega_1 \cap \Omega_2$, we have
\begin{equation*}
\begin{split}
\text{Cov}(\hat\eta_i, \hat\eta_j) &= \text{Cov}(\hat\eta_i^{(1)}, \lambda_j^{(1)} \hat\eta_j^{(1)} + \lambda_j^{(2)} \hat\eta_j^{(2)})\\
&= \lambda_j^{(1)} \text{Cov}(\hat\eta_i^{(1)}, \hat\eta_j^{(1)}) + \lambda_j^{(2)} \text{Cov}(\hat\eta_i^{(1)}, \hat\eta_j^{(2)})\\
&= -\frac{\lambda_j^{(1)} \eta_i \eta_j}{K_1}.
\end{split}
\end{equation*}
Similarly, we have
$\text{Cov}(\hat\eta_i, \hat\eta_j) = -\frac{\lambda_i^{(2)} \eta_i \eta_j}{K_2}$
when $i\in \Omega_1 \cap \Omega_2$ and $j\in \Omega_2 - \Omega_1$.
When $i, j\in \Omega_1 \cap \Omega_2$ and $i\ne j$, we have
\begin{equation*}
\begin{split}
\text{Cov}(\hat\eta_i, \hat\eta_j) &= \text{Cov}(\lambda_i^{(1)} \hat\eta_i^{(1)} + \lambda_i^{(2)} \hat\eta_i^{(2)}, \lambda_j^{(1)} \hat\eta_j^{(1)} + \lambda_j^{(2)} \hat\eta_j^{(2)})\\
&= \lambda_i^{(1)} \lambda_j^{(1)} \text{Cov}(\hat\eta_i^{(1)}, \hat\eta_j^{(1)}) + \lambda_i^{(2)} \lambda_j^{(2)} \text{Cov}(\hat\eta_i^{(2)}, \hat\eta_j^{(2)})\\
&= -\eta_i \eta_j\left(\frac{\lambda_i^{(1)} \lambda_j^{(1)}}{K_1} + \frac{\lambda_i^{(2)} \lambda_j^{(2)}}{K_2}\right).
\end{split}
\end{equation*}
Finally, the variance of $\hat \eta_2$ is computed as
\begin{equation*}
\begin{split}
&\text{Var}(\hat \eta_2) = \text{Var} (\Lambda_4 - \sum_{i\in \Omega_3^*} \phi_{i}^{(3)} \hat\eta_i)\\
&=\sum_{i\in \Omega_3^*} \text{Var} ( \phi_{i}^{(3)} \hat\eta_i) + \sum_{i\in \Omega_3^*} \sum_{j\ne i, j\in \Omega_3^*} \text{Cov}( \phi_{i}^{(3)} \hat\eta_i, \phi_{j}^{(3)} \hat\eta_j)\\
&=\sum_{i\in \Omega_3^*} (\phi_{i}^{(3)})^2 \text{Var} (\hat\eta_i) + \sum_{i\in \Omega_3^*} \sum_{j\ne i, j\in \Omega_3^*} \phi_{i}^{(3)} \phi_{j}^{(3)} \text{Cov}(\hat\eta_i, \hat\eta_j).
\end{split}
\end{equation*}

\balance
\bibliographystyle{abbrv}

\begin{thebibliography}{10}

\bibitem{AhmedKDD2014}
N.~Ahmed, N.~Duffield, J.~Neville, and R.~Kompella.
\newblock Graph sample and hold: A framework for big-graph analytics.
\newblock In {\em Proceedings of the 20th ACM SIGKDD International Conference
  on Knowledge Discovery and Data Mining}, pages 589--597, 2014.

\bibitem{Albert2004}
I.~Albert and R.~Albert.
\newblock Conserved network motifs allow protein--protein interaction
  prediction.
\newblock {\em Bioinformatics}, 4863(13):3346--3352, 2004.

\bibitem{Alon1995}
N.~Alon, R.~Yuster, and U.~Zwick.
\newblock Color-coding.
\newblock {\em J. ACM}, 42(4):844--856, July 1995.

\bibitem{Bhuiyan2012}
M.~A. Bhuiyan, M.~Rahman, M.~Rahman, and M.~A. Hasan.
\newblock Guise: Uniform sampling of graphlets for large graph analysis.
\newblock In {\em Proceedings of IEEE ICDM 2012}, pages 91--100, December 2012.

\bibitem{ChengLLFLH15}
J.~Cheng, Q.~Liu, Z.~Li, W.~Fan, J.~C.~S. Lui, and C.~He.
\newblock {VENUS:} vertex-centric streamlined graph computation on a single
  {PC}.
\newblock In {\em 31st {IEEE} International Conference on Data Engineering,
  {ICDE} 2015, Seoul, South Korea, April 13-17, 2015}, pages 1131--1142, 2015.

\bibitem{ChunIMC2008}
H.~Chun, Y.~yeol Ahn, H.~Kwak, S.~Moon, Y.~ho~Eom, and H.~Jeong.
\newblock Comparison of online social relations in terms of volume vs.
  interaction: A case study of cyworld.
\newblock In {\em Proceedings of ACM SIGCOMM Internet Measurement Conference
  2008}, pages 57--59, November 2008.

\bibitem{Hasan2009}
M.~A. Hasan and M.~J. Zaki.
\newblock Output space sampling for graph patterns.
\newblock In {\em Proceedings of the VLDB Endowment 2009}, pages 730--741,
  August 2009.

\bibitem{IliofotouCoNEXT2009}
M.~Iliofotou, M.~Faloutsos, and M.~Mitzenmacher.
\newblock Exploiting dynamicity in graph-based traffic analysis: Techniques and
  applications.
\newblock In {\em Proceedings of the 5th International Conference on Emerging
  Networking Experiments and Technologies}, CoNEXT 2009, pages 241--252, 2009.

\bibitem{Itzkovitz2005}
S.~Itzkovitz, R.~Levitt, N.~Kashtan, R.~Milo, M.~Itzkovitz, and U.~Alon.
\newblock Coarse-graining and self-dissimilarity of complex networks.
\newblock {\em Physica Rev.E}, 71:016127, 2005.

\bibitem{JhaKDD2013}
M.~Jha, C.~Seshadhri, and A.~Pinar.
\newblock A space efficient streaming algorithm for triangle counting using the
  birthday paradox.
\newblock In {\em Proceedings of the 19th ACM SIGKDD International Conference
  on Knowledge Discovery and Data Mining}, pages 589--597, 2013.

\bibitem{JhaWWW15}
M.~Jha, C.~Seshadhri, and A.~Pinar.
\newblock Path sampling: {A} fast and provable method for estimating 4-vertex
  subgraph counts.
\newblock In {\em Proceedings of the 24th International Conference on World
  Wide Web, {WWW} 2015, Florence, Italy, May 18-22, 2015}, pages 495--505,
  2015.

\bibitem{JhaKDD2015}
M.~Jha, C.~Seshadhri, and A.~Pinar.
\newblock Path sampling: {A} fast and provable method for estimating 4-vertex
  subgraph counts.
\newblock In {\em Proceedings of the 24th International Conference on World
  Wide Web, {WWW} 2015, Florence, Italy, May 18-22, 2015}, pages 495--505,
  2015.

\bibitem{JinSigmetric2009}
Y.~Jin, E.~Sharafuddin, and Z.-L. Zhang.
\newblock Unveiling core network-wide communication patterns through
  application traffic activity graph decomposition.
\newblock In {\em Proceedings of the Eleventh International Joint Conference on
  Measurement and Modeling of Computer Systems}, SIGMETRICS 2009, pages 49--60,
  2009.

\bibitem{Kashtan2004}
N.~Kashtan, S.~Itzkovitz, R.~Milo, and U.~Alon.
\newblock Efficient sampling algorithm for estimating subgraph concentrations
  and detecting network motifs.
\newblock {\em Bioinformatics}, 20(11):1746--1758, 2004.

\bibitem{Kunegis2009}
J.~Kunegis, A.~Lommatzsch, and C.~Bauckhage.
\newblock The slashdot zoo: mining a social network with negative edges.
\newblock In {\em Proceedings of WWW 2009}, pages 741--750, April 2009.

\bibitem{KyrolaOSDI12}
A.~Kyrola, G.~E. Blelloch, and C.~Guestrin.
\newblock Graphchi: Large-scale graph computation on just a {PC}.
\newblock In {\em 10th {USENIX} Symposium on Operating Systems Design and
  Implementation, {OSDI} 2012, Hollywood, CA, USA, October 8-10, 2012}, pages
  31--46, 2012.

\bibitem{LeskovecTKDD2007}
J.~Leskovec, J.~Kleinberg, and C.~Faloutsos.
\newblock Graph evolution: Densification and shrinking diameters.
\newblock {\em Transactions on Knowledge Discovery from Data (TKDD)}, 1(1),
  Mar. 2007.

\bibitem{LeskovecIM2009}
J.~Leskovec, K.~J. Lang, A.~Dasgupta, and M.~W. Mahoney.
\newblock Community structure in large networks: Natural cluster sizes and the
  absence of large well-defined clusters.
\newblock {\em Internet Mathematics}, 6(1):29--123, 2009.

\bibitem{LowPVLDB2012}
Y.~Low, J.~Gonzalez, A.~Kyrola, D.~Bickson, C.~Guestrin, and J.~M. Hellerstein.
\newblock Distributed graphlab: A framework for machine learning in the cloud.
\newblock {\em PVLDB}, 5(8):716--727, 2012.

\bibitem{Pregel2010}
G.~Malewicz, M.~H. Austern, A.~J. Bik, J.~C. Dehnert, I.~Horn, N.~Leiser, and
  G.~Czajkowski.
\newblock Pregel: A system for large-scale graph processing.
\newblock In {\em Proceedings of the 2010 ACM SIGMOD International Conference
  on Management of Data}, pages 135--146, 2010.

\bibitem{OmidiGenes2009}
S.~Omidi, F.~Schreiber, and A.~Masoudi-nejad.
\newblock Moda: An efficient algorithm for network motif discovery in
  biological networks.
\newblock {\em Genes and Genet systems}, 84(5):385--395, 2009.

\bibitem{PavanyVLDB2013}
A.~Pavany, K.~T.~S. Tirthapuraz, and K.-L. Wu.
\newblock Counting and sampling triangles from a graph stream.
\newblock In {\em Proceedings of VLDB}, pages 1870--1881, 2013.

\bibitem{GRAFT2012}
M.~Rahman, M.~Bhuiyan, and M.~A. Hasan.
\newblock Graft: An approximate graphlet counting algorithm for large graph
  analysis.
\newblock In {\em Proceedings of the 21st ACM International Conference on
  Information and Knowledge Management}, 2012.

\bibitem{Richardson2003}
M.~Richardson, R.~Agrawal, and P.~Domingos.
\newblock Trust management for the semantic web.
\newblock In {\em Proceedings of the 2nd International Semantic Web
  Conference}, pages 351--368, October 2003.

\bibitem{Ripeanu2002}
M.~Ripeanu, I.~T. Foster, and A.~Iamnitchi.
\newblock Mapping the gnutella network: Properties of large-scale peer-to-peer
  systems and implications for system design.
\newblock {\em IEEE Internet Computing Journal}, 6(1):50--57, 2002.

\bibitem{Shenorr2002}
S.~S. Shen-Orr, R.~Milo, S.~Mangan, and U.~Alon.
\newblock Network motifs in the transcriptional regulation network of
  escherichia coli.
\newblock {\em Nature Genetics}, 31(1):64--68, May 2002.

\bibitem{TsourakakisKDD2009}
C.~E. Tsourakakis, U.~Kang, G.~L. Miller, and C.~Faloutsos.
\newblock Doulion: Counting triangles in massive graphs with a coin.
\newblock In {\em PROCEEDINGS OF ACM KDD 2009}, 2009.

\bibitem{Ugander2013}
J.~Ugander, L.~Backstrom, and J.~Kleinberg.
\newblock Subgraph frequencies: mapping the empirical and extremal geography of
  large graph collections.
\newblock In {\em Proceedings of the 22nd international conference on World
  Wide Web}, WWW 2013, pages 1307--1318, 2013.

\bibitem{TKDDWang2014}
P.~Wang, J.~C. Lui, J.~Zhao, B.~Ribeiro, D.~Towsley, and X.~Guan.
\newblock Efficiently estimating motif statistics of large networks.
\newblock {\em ACM Transactions on Knowledge Discovery from Data}, 2014.

\bibitem{PinghuiMotifEdgeSampling2014}
P.~Wang, J.~C.~S. Lui, and D.~Towsley.
\newblock Minfer: Inferring motif statistics from sampled edges.
\newblock {\em CoRR}, abs/1502.06671, 2015.

\bibitem{Wernicke2006}
S.~Wernicke.
\newblock Efficient detection of network motifs.
\newblock {\em IEEE/ACM Transactions on Computational Biology and
  Bioinformatics}, 3(4):347--359, 2006.

\bibitem{YangICDM2012}
J.~Yang and J.~Leskovec.
\newblock Defining and evaluating network communities based on ground-truth.
\newblock In {\em 12th IEEE International Conference on Data Mining (ICDM)},
  pages 745--754, 2012.

\bibitem{ZhaoNetsci2011}
J.~Zhao, J.~C.~S. Lui, D.~Towsley, X.~Guan, and Y.~Zhou.
\newblock Empirical analysis of the evolution of follower network: A case study
  on douban.
\newblock In {\em Proceedings of IEEE INFOCOM NetSciCom 2011}, pages 941--946,
  April 2011.

\end{thebibliography}

\end{document}